\documentclass[aps,prf,amsmath,amssymb,groupedaddress]{revtex4-2}

\usepackage{graphicx}
\usepackage{dcolumn}
\usepackage{bm}
\usepackage[normalem]{ulem}
\usepackage{xcolor}

\usepackage[utf8]{inputenc}
\usepackage[T1]{fontenc}
\usepackage{mathptmx}
\usepackage{etoolbox}

\usepackage{subeqnarray}
\usepackage{amsmath}
\usepackage{float}
\usepackage{amsfonts}
\usepackage{epsfig}

\makeatletter
\def\@email#1#2{%
	\endgroup
	\patchcmd{\titleblock@produce}
	{\frontmatter@RRAPformat}
	{\frontmatter@RRAPformat{\produce@RRAP{*#1\href{mailto:#2}{#2}}}\frontmatter@RRAPformat}
	{}{}
}%
\makeatother
\begin{document}


\title{Fore-aft clearance controls how three-dimensional confinement affects micropropulsion}


\author{Suraj Kumar Kamarapu}
\affiliation{ 
	Department of Mechanical Engineering, University of Utah, Salt Lake City, UT 84112 USA.
}
\author{Mehdi Jabbarzadeh}%
\affiliation{ 
	Department of Mechanical Engineering, University of Utah, Salt Lake City, UT 84112 USA.
}
\author{Henry Chien Fu*}
\email{henry.fu@utah.edu}
\affiliation{ 
	Department of Mechanical Engineering, University of Utah, Salt Lake City, UT 84112 USA.
}%
\homepage{http://hfu.mech.utah.edu}



\date{\today}

\begin{abstract}
Systems of active particles are often affected by confinement due to nearby boundaries.  Recently, there has been interest in the effect of confinement by complex three dimensional geometries, as might occur in structured environments such as porous media, foams, gels, or biological tissues and ducts.  The effects of confinement for particles moving along boundaries has been extensively studied, but in three dimensions active particles move not only parallel to boundaries, but also towards or away from boundaries.  The consequences of this fore-aft clearance is less well understood.  Swimmers that actively remodel their environment create an ideal situation to study the effect of clearance, since they maintain a steady clearance while translating. By numerically studying the locomotion of the bacterium \textit{Helicobacter pylori}, which de-gels surrounding gastric mucus to make a co-moving pocket of fluid around itself, we show that the effect of three-dimensional confinement  is controlled by clearance, rather than distance from a parallel boundary. Analytical calculations show that the effect of clearance can be understood in terms of flow structures, such as the generic pusher and puller flows of active particles, indicating that our results should apply to a wide range of confined active particles. 
\end{abstract}

\maketitle


Microscale active self-propelled particles \cite{Bechinger2016}, including microbial swimmers \cite{Lauga2009,Spagnolie2012,Elgeti2015,Berg1973} and engineered colloidal particles \cite{Walther2013,Sanchez2012}, are of current interest for their biological \cite{Chen2020, Beer2019,Schwarz-Linek2012} and ecological \cite{Michalec2017, Gozdziejewska2019, Fenchel1986, Kiorboe2016, Azam1994} importance, for biomedical and technological applications \cite{Xu2018, Stenhammar2015,Kim2013}, and as key components of nonequilibrium states of matter \cite{Bechinger2016}.
Active particles often move through complex media that are not simple Newtonian fluids, both in natural \cite{Friedrich2007,Kirkman-Brown2011,Tung2017, Harman2012,Figueroa-Morales2019,Fu2007a, Jabbarzadeh2014} and  laboratory \cite{Martinez2014,Wang2019, Guillamat2017, Aranson2018, Juarez2010} environments, including many situations where geometrical confinement becomes important.  For example, bacteria can have altered trajectories \cite{Lauga2006, Molaei2016} and accumulate when swimming along boundaries \cite{Berke2008,Li2009, Deng2020, Hu2015}, including in naturally arising situations such as oil drops \cite{Desai2018,Prakash2020,Fernandez2019}, while  engineered boundaries can be designed with complex shapes to guide and collect bacteria \cite{Galajda2007,Austin2017,Bianchi2017} and sperm \cite{Secchi2020, Denissenko2012, Kantsler2014,Rode2019} along them.

\begin{figure}[t]
	\begin{center}
		\includegraphics[width=0.4\columnwidth]{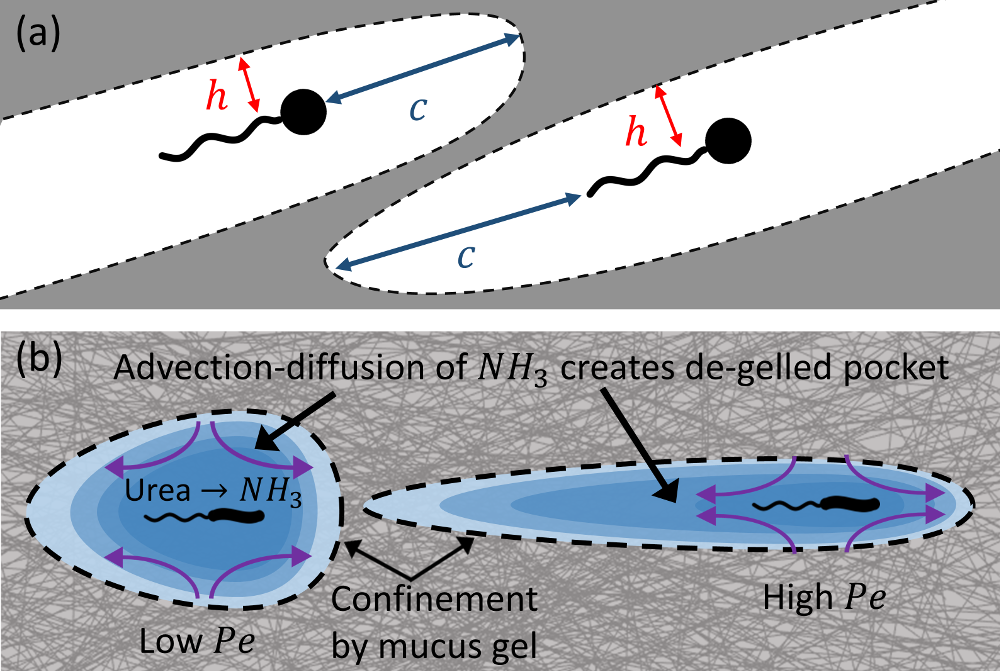}
	\end{center}
	\caption{a) Bacteria swimming in 3D confinement with parallel distance $h$ and fore-aft clearance $c$ to boundaries.
		b) Active remodeling creates steady fore-aft clearance; \textit{e.g.}, ammonia diffusing from \textit{H. pylori} dissolves mucus gel to create a fluid pocket.  An almost spherical pocket forms at low $Pe$ (diffusion dominated) while a slender pocket forms at high $Pe$ (advection dominated).}
	\label{fig:schematic} 
\end{figure}

In three dimensions, confinement is not restricted to swimming parallel to a boundary (characterized by the distance $h$ in Fig.~\ref{fig:schematic}a), but can also involve fore-aft clearance in front or behind the swimmer ($c$ in Fig.~\ref{fig:schematic}a), such as in porous media \cite{Kuhn2017, Bhattacharjee2019}, sea foams \cite{Jenkinson2018}, and microfluidic devices \cite{Majmudar2012, Li2014, Dehkharghani2019}.
Since the clearance often changes as a swimmer translates, it can lead to complex dynamics \cite{Narinder2021, Ishimoto2020} and be difficult to study. However, steady time-independent clearance can be realized by swimmers that actively remodel their environment, making them ideally suited to investigate three-dimensional confinement. For example, when swimming through gastric mucus gel, the bacterium \textit{Helicobacter pylori} actively degels surrounding mucus by producing neutralizing ammonia \cite{Celli2009}, creating a pocket that moves with the bacterium of fluid confined by gel (Fig.~\ref{fig:schematic}b). Local modification of material parameters also occurs chemically near microrobots \cite{Walker2015a}, in the phycospheres of plankton \cite{Guadayol2020, Shoele2018a, Eastham2020a, Dandekar2020}, and by thermal \cite{Jiang2010, Rings2010} and mechanical effects \cite{Martinez2014, Zottl2019}.  

Despite its importance, the effect of three-dimensional clearance is not as well understood as parallel confinement, which has been analyzed from many perspectives \cite{Spagnolie2012, Shen2017, Kurzthaler2021}. 
Here, by studying the locomotion of actively-remodeling \textit{H. pylori}, we show that in truly three dimensional confinement swimming behavior is most strongly affected not by the distance from a parallel boundary ($h$), but by the less-appreciated fore-aft clearance of the swimmer. We can understand the effects of clearance in terms of far-field pusher- and puller-type hydrodynamic flows, indicating that our results should be generally applicable, including for active particles in three-dimensional confinement which is not self-created, and hence has time-varying clearance.

\section{Active remodeling of confinement by \textit{H. pylori}}
We investigate three dimensional confinement in the context of \textit{H. pylori}, which infects the stomach of roughly half the world's humans, causing gastritis, peptic ulcers, and gastric cancer \cite{Zamani2018}.
\textit{H. pylori} has evolved to survive in the human stomach, despite the acidic environment (pH $<$ 2) in its lumen required for food digestion. 
A gastric mucus layer protects the stomach epithelium (pH $\approx$ 6-7) from the acidic lumen (pH $<$ 2) and its digestive proteins. \textit{H. pylori} must swim across this barrier to colonize the epithelium.
To do so, it hydrolyzes ambient urea to produce basic ammonia, which turns gastric mucus from a gel (pH $<$ 4) to a fluid (pH $>$ 4) \cite{Celli2007} in which it can swim \cite{Celli2009}. 
For isolated \textit{H. pylori}, ammonia, neutralization, and degelation are localized, forming a pocket of fluid moving with the bacterium \cite{Mirbagheri2016} as it swims (Fig. \ref{fig:schematic}b).
\textit{H. pylori} therefore actively remodels its swimming environment via a process coupling geometrical effects and chemical diffusion --  
the pocket geometry affects swimming and flows through confinement effects, while the flows simultaneously alter the spread of ammonia, hence neutralization and pocket geometry, through advection-diffusion.

It remains unclear how swimming in a co-moving pocket of confinement affects swimming speeds, with contradictory results from previous theoretical studies. 
A model of pocket formation \cite{Mirbagheri2016} simplified the confinement problem to a two-dimensional waving sheet swimming in a layer of fluid of thickness $h$ confined by a Brinkman medium representing the mucus gel \cite{Leshansky2009}, resulting in increased swimming speeds for all $h$. In contrast, fully three-dimensional analytic models \cite{Reigh2017, Nganguia2020} of a spherical squirmer in a spherical pocket of fluid confined by another fluid of higher viscosity \cite{Reigh2017}, or a Brinkman medium \cite{Nganguia2020}, found that squirmers with prescribed tangential surface velocities slowed down, but those with radial surface velocities sped up.

\label{methods}

Previous investigations were also restricted to the bacteria-appropriate low Peclet number ($Pe$) regime, in which diffusion dominates advection, leading to a relatively large and nearly spherical pocket (Fig.~\ref{fig:schematic}b). However, microrobots which mimic \textit{H. pylori's} strategy to swim through mucus \cite{Walker2015} closely approach the front of a non-spherical pocket to bore through the gel (Fig.~\ref{fig:schematic}b), seemingly in the high Peclet number regime dominated by advection. To clarify how three-dimensional confinement affects swimmers across the full range of Peclet numbers, we implement a fully three-dimensional numerical treatment of active remodeling of the environment coupled to swimming locomotion that can accurately treat realistic bacterial and confinement geometries.

\section{Hydrodynamic model}
In the low-Reynolds number regime dominated by viscosity $\kappa$ which is appropriate at microscales, the swimming speed is determined by solving for an incompressible velocity $\mathbf{v}$ and pressure $p$ field which satisfy Stokes equation,
\begin{equation}
0 = -\nabla p + \kappa \nabla^2 \mathbf{v},
\end{equation}
and the no-slip boundary conditions on a bacterium with translational and angular velocities $\mathbf{V}_c$ and $\bm{\Omega}_c$ and a prescribed relative rotation between cell body and flagellum $\bm{\Omega}_{rel}$, while exerting no net force or torque on the bacterium.
We obtain the solution using the method of regularized Stokeslets \cite{Cortez2002,Cortez2005,Hyon2012,Martindale2016}, in which the velocity field is represented as a superposition of contributions from regularized Stokeslet solutions resulting from a force acting on a `blob' of fluid.  \textit{H. pylori} is modeled using a spherical cell body with attached helical flagellar bundle (Fig.~\ref{sphereconfine}a), both discretized on their surfaces with a distribution of regularized Stokeslets as in \cite{Martindale2016}.  
The confining mucus gel is modeled by a layer of randomly distributed static regularized Stokeslets ($\mathbf{v}=0$ at Stokeslet locations).  This is similar to previously employed models \cite{Leshansky2009} and the density and size of the regularized Stokeslets can be chosen to correspond to a Brinkman medium \cite{Kamarapu2021}.  
Modeling confinement typically involves ${\cal O}(10^6)$ regularized Stokeslets, and we use the Kernel Independent Fast Multipole Method \cite{Rostami2016,Rostami2019} to solve the system of equations that result from velocity conditions specified at each Stokeslet and the force- and torque free constraints.
Details of geometries, discretization, and numerical solution are in Appendix \ref{HDmodel}. 

\noindent\textbf{Coupling confinement hydrodynamics and gel remodeling.}\label{fpg}
To determine a pocket geometry, the flow field obtained by the method of regularized Stokeslets is used to specify an advection diffusion problem for the ammonia concentration ($C$) around the swimmer, assuming that the bacteria self-regulates a concentration of ammonia at its surface corresponding to pH=6. The pocket boundary is the surface with a hundredfold decrease of ammonia concentration, hence pH=4, the critical degelation pH \cite{Mirbagheri2016}.  


In the lab frame, bacterial translation and rotation leads to unsteady flow fields ($\mathbf{v}$) and concentrations. However, for an axisymmetric cell body the problem can be greatly simplified by working in the frame of the translating and rotating flagellum, in which the flow field is steady and we find the ammonium concentration $C$, defined using the position vector from the reference point on flagellum, satisfying the steady-state advection diffusion equation,
\begin{equation}
0 =
\left(\textbf{v} - \textbf{V} - \bm{\Omega}\times r\right)\cdot \nabla C - D\nabla^2 C,
\end{equation}
which we solve using the COMSOL package
(details in Appendices \ref{A-D} and \ref{Iterative}).

The resulting pocket geometry may not be the same as the geometry that produced the hydrodynamic flows used in the advection-diffusion equation.
Therefore we iterate, solving for hydrodynamic flows using the new pocket geometry, then finding the next pocket geometry using those hydrodynamic flows, and repeating until we find a self-consistent solution for the pocket geometry that simultaneously produces and results from the hydrodynamic flows.  We stop iterating when the pocket geometries and swimming speeds between two successive iterations change less than $5\%$.

\begin{figure}[]
	\includegraphics[width=0.5\columnwidth]{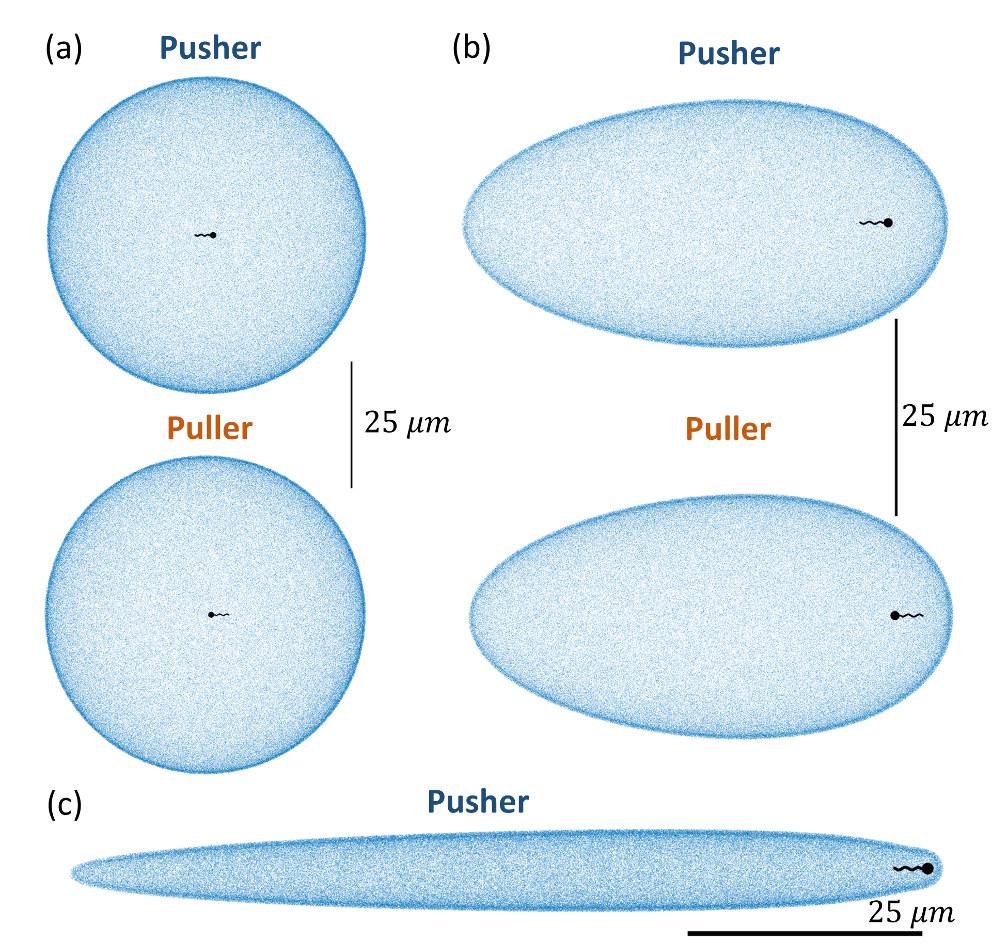} 
	\caption{Bacterium (swimming right) in self-created fluid pocket at (a) low $Pe = 1.52 \times 10^{-3}$, (b) intermediate $Pe = 1.52 \times 10^{-1}$, and (c) high $Pe=6$. Blue points show the shell of regularized Stokeslets confining the pocket.}
	\label{fluidpocket}
\end{figure}

\begin{figure}[t]
	\includegraphics[width=0.5\columnwidth]{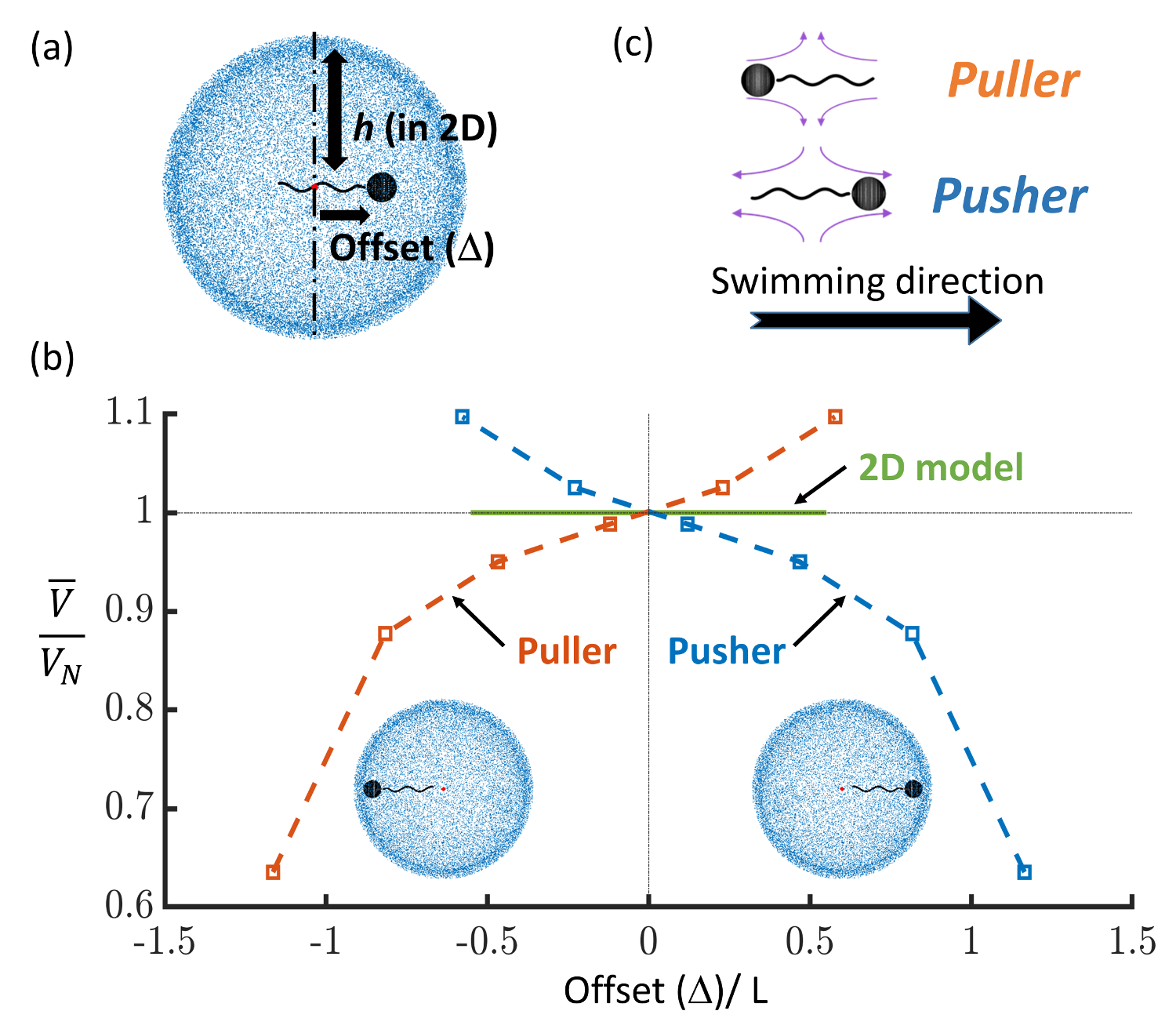} 
	\caption{(a) A swimmer in spherical confinement with offset $\Delta$ in the swimming direction.  (b) Unlike in 2D (green) \cite{Mirbagheri2016}, 3D confinement effects on swimming speed $\overline{V}$ (normalized by unconfined speed $V_N$) are larger, can slow down as well as speed up, and are controlled by offset (normalized by flagellar length $L$).
		(c) Flows around puller and pushers interact with boundaries to affect $\overline{V}$.}
	\label{sphereconfine}
\end{figure}

We obtained self-consistent solutions for both pusher and puller bacteria in diffusion-dominated (small $Pe$), advection-dominated (large $Pe$) and intermediate (medium $Pe$) regimes. 
A typical Peclet number for \textit{H. pylori} is $ Pe = \overline{V}d/(2D) = 1.5 \times 10^{-3}$, using thickness $d=0.5\,\mu$m, average swimming speed  $ \overline{V} = 10 \, \mu$m/s \cite{Martinez2016}, and the diffusion constant of ammonia in water $D = 1.64\times10^{-9} \mathrm{m}^2/s$ \cite{Cussler}. 
At this low Peclet number, ammonia diffuses far from the cell body and produces a large, nearly spherical pocket (Fig.~\ref{fluidpocket}a).  The flow is barely affected by the distant confinement; it changes the swimming speed by less than $0.02\%$.
At intermediate $Pe$= 0.15, the importance of advection increases and ammonia and the pocket trail behind the moving swimmer, which ends up offset further forward (Fig.~\ref{fluidpocket}b).  
At large $Pe$=6, as might occur for a fast-moving microrobot, the pocket is stretched out and the swimmer shifted forward even more (Fig.~\ref{fluidpocket}c). If the swimmer is close to the pocket boundary, the confinement geometry can be strongly affected by details of the near-field swimmer flow where the boundary is moved outward by the pusher flow. 
We do not treat a puller at large Peclet number, since the flagellum may directly contact the pocket boundary. Contact with the confinement would require a model that incorporates direct gel interactions and deformation \cite{Fu2010a}, and we do not attempt to describe such a scenario here.

%
%

For the pusher, relative to the unconfined speed $V_N$, the swimming speed $\overline{V}$ is reduced to $0.769 V_N$, $0.987 V_N$, and $0.99994 V_N$ at $Pe$= 6, 0.15, and 0.0015, respectively, while for the puller $\overline{V}$ increases to $1.02 V_N$ and $1.0002 V_N$ at $Pe$=0.15 and 0.0015, respectively. 
Interestingly, the observed decrease in swimming speed for pushers is opposite of the effect of 2D confinement \cite{Mirbagheri2016} on swimming sheets, and that of spherically symmetric confinement of squirmers propelled by radial deformations \cite{Nganguia2020}.

\section{3D confinement effect is controlled by clearance}\label{3dconfineT}
To explain these results, we explore the effect of confinement on a bacterium in a spherical pocket.  In particular, fore-aft clearance is included via the swimmer's offset $\Delta$ from the center of the sphere (Fig.~\ref{sphereconfine}a).  
Offset (hence clearance) determines whether swimming slows down or speeds up.
The swimming speed ($\overline{V}$) of a puller bacterium (body behind the flagellum) is strongly dependent on the offset (Fig.~\ref{sphereconfine}b); it is faster than the unconfined speed ($V_N$) when the bacterium is shifted forward in the pocket ($\Delta > 0$), and slower when the bacterium is shifted backward.  
Kinematic reversibility \cite{Purcell1977} implies that a pusher bacterium (flagellum behind the body) has equal but opposite swimming velocity as a puller bacterium in the same configuration. Because offset is defined relative to the swimming direction, a pusher that is shifted backward ($\Delta < 0$) speeds up, as confirmed by direct calculations for pusher bacteria (see Appendix \ref{HDmodel}).
Note that we chose the offset (defined relative to an arbitrary point on the swimmer) so that $\Delta=0$ corresponds to $\overline{V}=V_N$.  

The differing effect of offset on pushers and pullers can be understood in terms of their respective flow fields (Fig.~\ref{sphereconfine}c).  A pusher has outward flow fore-and-aft which pushes away from the closest boundary, speeding it up when near the back surface, and slowing it down near the front surface. The puller has inward flow fore-and-aft that produces the opposite effect. To confirm this flow-based picture, we calculated the speed of a spherical squirmer propelled by prescribed tangential slip velocities in a spherical pocket with nonzero offset.  
In a bispherical coordinate system $\lbrace\xi, \eta, \phi \rbrace$, both the surface of the squirmer and the pocket can be chosen to be surfaces of constant $\xi$, which allows the solution to be found by separation of variables \cite{Jabbarzadeh2018} (Fig. \ref{squirmer}a, details in Appendix \ref{Bispherical}).  Behind a pusher-type squirmer, in the far field the flow moves away from the swimmer and decays as $1/r^2$, but closer to the swimmer there is a stagnation point at a distance $d_{stag}$ from the squirmer surface,  and at even closer locations the flow reverses to satisfy the  boundary conditions at the forward-moving swimmer surface (Fig. \ref{squirmer}b).  

To see how this flow impacts the effect of clearance on swimming, we calculate the swimming speed as a function of clearance $c$ (Fig. \ref{squirmer}a) from the rear boundary of the spherical pocket (Fig.~\ref{squirmer}c).
As the clearance $c$ from the rear pocket surface decreases, at first the swimming speed increases since the backwards flow velocity increases.  However, as the clearance moves through the stagnation point and the flow switches directions, the swimming speed reaches a maximum and then decreases, even slowing below the unconfined $V_N$ for the smallest clearances. 
This behavior confirms that confinement effects are caused by flow pushing or pulling on the boundary.
Indeed, the clearance ($c_{max}$) at which the swimming speed enhancement is maximum has a linear correlation with the position of the stagnation point ($d_{stag}$) as the squirmer stroke varies (Fig. \ref{squirmer}c inset).

\begin{figure}
	\includegraphics[width=0.5\columnwidth]{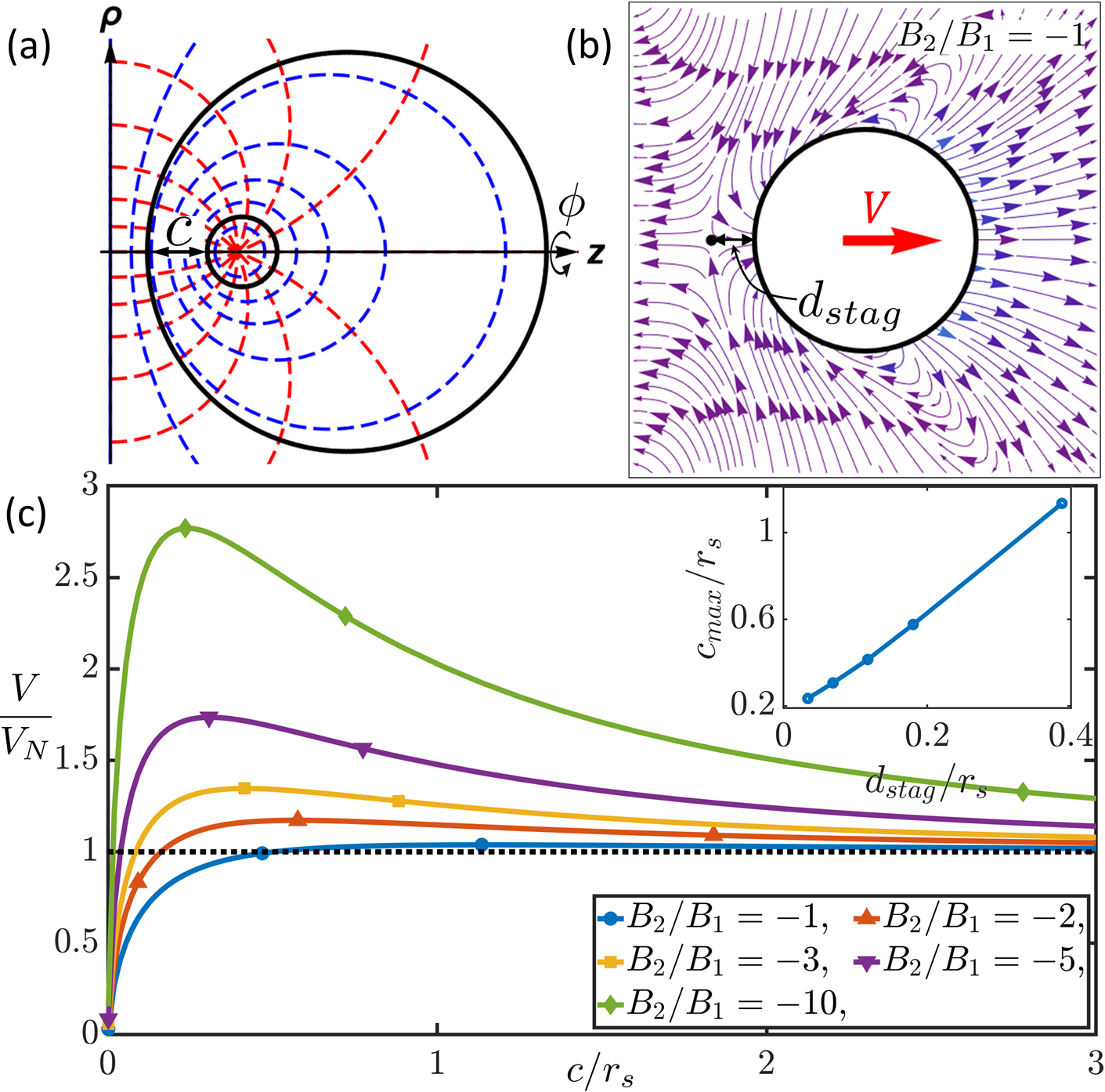} 
	\caption{(a) Surfaces (black) of a spherical pocket and squirmer (offset with clearance $c$) are both surfaces of constant $\xi$ (blue) in the bispherical coordinate system ($\xi$,$\eta$,$\phi$).  Surfaces of constant $\eta$ are red. 
		(b) Streamlines showing that behind (to left of) an isolated pusher squirmer there is backwards far field flow, a stagnation point at distance $d_{stag}$, and flow reversal near the squirmer surface. 
		(c) Swimming speed $V$ as a function of clearance reflects the flow reversal since the flow pushes and pulls on the pocket wall for large and small clearance, respectively. Clearance $c_{max}$ with maximum speed enhancement linearly correlates with $d_{stag}$ for various swimming strokes specified by parameters $B_1$ and $B_2$ (see SM).}
	\label{squirmer}
\end{figure}

\section{Discussion}
We have shown that the effect of the three dimensional confinement created by \textit{H. pylori} is controlled by the swimmer's offset within the pocket, hence fore-aft clearance. 
In an actively remodeled pocket, pusher and puller bacteria speed up and slow down, respectively, consistent with the clearance dependence elucidated using a spherical pocket, since in both cases the bacterium is offset forward in the pocket due to its translation.  
The change in swimming speed is larger at high Peclet numbers due to the larger forward offset and proximity to the pocket boundary.
Previous studies could not access this clearance effect since they used either two-dimensional models of confinement, or spherically symmetric geometries with no offset. 
Our results encompass both low- and high-Peclet number regimes, thus providing insight into the full range of behavior for active particles -- both biological and engineered -- that remodel their environment. Artificial microswimmers that traverse the gastric mucus barrier, such as for drug delivery, employ a similar strategy as $\textit{H. pylori}$ \cite{Walker2015a}.  Indeed, the microswimmers of \cite{Walker2015a} are expected to have a pusher-type flow, and their large forward offset and slow speeds are consistent with our results.

We expect that the described clearance dependence is generic for dipolar swimmers in three dimensional confinement, since the effect of the offset for pushers and pullers can be rationalized in terms of their hydrodynamic flows. However, as suggested by our calculation for squirmer swimmers, confinement effects should depend on the details of geometries and swimming strokes if the swimmer closely approaches the boundary and near-field hydrodynamics become relevant.

Active remodeling is an ideal scenario to study three-dimensional confinement since it results in a time-independent fore-aft clearance. Active remodeling occurs quite frequently if one expands the notion of confinement to include not only nearby solid or gel boundaries, but also heterogeneous viscoelastic or viscous resistance.  As examples, in polymeric solutions swimmers can produce nearby flows enhancing shear-thinning \cite{Qu2020, Li2015, Gomez2017, Pietrzyk2019}, and rotating bacterial flagella can even deplete nearby polymeric concentrations \cite{Martinez2014, Zottl2019}.  In yield-stress materials a swimmer also can create a nearby fluidized region \cite{Hewitt2017}, and swimmers moving across an interface of fluids with different viscosities may drag a surrounding amount of different-viscosity fluid with them \cite{EsparzaLopez2021}. 
Active remodeling of local material properties can also be caused by thermal or chemical effects; plankton and bacteria can exude or consume viscosity-altering substances \cite{Guadayol2020, Shoele2018a, Eastham2020a, Dandekar2020}, thermal heating of microrobots can locally decrease viscosity, and the degelling strategy of \textit{H. pylori} has been mimicked in magnetically rotated microrobots \cite{Walker2015a}. 

Finally, the effect of fore-aft clearance is important in  complex geometries that are not self-generated, such as in structured environments like porous media and foams. Our results allow a physical understanding of these scenarios, as well as a route to understanding transport in such environments by including the dynamic clearance resulting from swimmer translation.

\noindent\textbf{Acknowledgments.} This work was supported by NIH award 1R01GM131408, and the Utah Center for High Performance Computing.



\appendix
\section{Solution of hydrodynamic model} \label{HDmodel}

In the method of regularized Stokeslets \cite{Cortez2002,Cortez2005}, the Stokes equation is satisfied by a hydrodynamic velocity field $\textbf{v}(\textbf{r})$ at a position $\textbf{r}$ consisting of the superposition of flows generated by regularized local forces. In our problem there are $N_s$ forces ($\textbf{f}^{\alpha}$) acting at locations $\textbf{r}^\alpha$ on the swimmer surface, with $1 \leq\alpha\leq N_s$, and $N_p$ forces  acting at locations ($\textbf{r}^\alpha$) within the porous medium, with $N_s+1 \leq\alpha\leq N_s + N_p$, for a total of $N=N_s +N_p$ regularized Stokeslets. Thus,
\begin{equation}\label{velocity}
v_i(\textbf{r}) = \sum_{\alpha=1}^{N} S_{ij}(\textbf{r} - \textbf{r}^\alpha) f_j^\alpha ,
\end{equation} 
where $S_{ij}(\textbf{r}, \epsilon) = [(r^2 + 2\epsilon^2)\delta_{ij} + r_ir_j]/[8\pi\kappa(r^2 + \epsilon^2)^{3/2}]$ is the regularized Stokeslet \cite{Cortez2005}, $\kappa$ is the viscosity of the fluid, and $\epsilon$ is the width of a regularized Stokeslet's force distribution. In all equations, we use indicial notation with subscript indices for cartesian coordinates, and repeated indices are implicitly summed over.

\begin{figure}[]
	\includegraphics[width=0.4\textwidth]{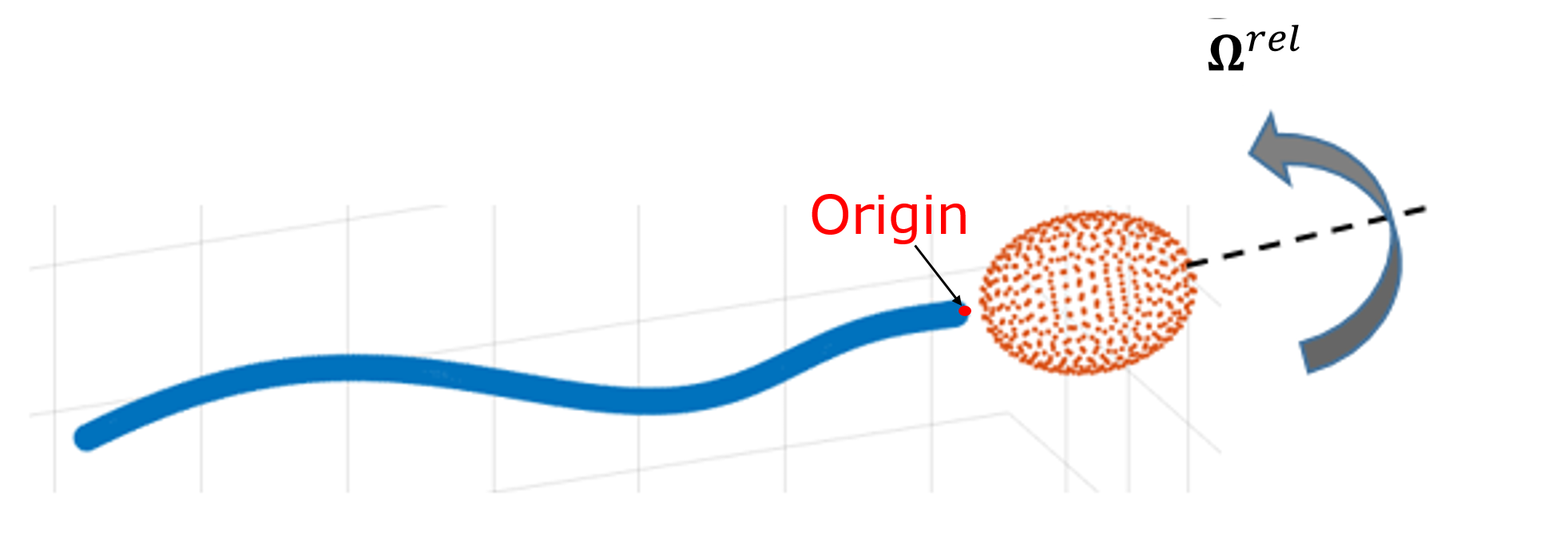} 
	\caption{Schematic of a bacterium swimming with a relative rate of rotation between cell body and flagellum, $\bm{\Omega}^{rel}$. }
	\label{orel}
\end{figure}

The fluid velocity satisfies the following boundary conditions.  The regularized Stokeslets representing the porous medium are static, so at their positions the velocity is zero,
\begin{equation}
\mathbf{v}(\mathbf{r}^\alpha)=0 \label{constraints3}, \qquad\qquad N_s+1 \leq \alpha \leq N.\\
\end{equation} 
We choose to work in body frame coordinates, where all positions are specified by their displacement $\mathbf{r}$ from the origin defined by the attachment point of the flagellum on the cell body (Fig. ~\ref{orel}). Furthermore, we work in a frame that co-rotates with the flagellum, so motor rotation means that the head rotates with (prescribed) angular velocity $\bm{\Omega}^{rel}$ relative to the flagellum. We denote the translational and angular velocities of the origin in the lab frame with $\mathbf{V}$ and $\bm{\Omega}$, respectively. Therefore the no-slip boundary conditions at the surface of the swimmer require that
\begin{equation}
\mathbf{v}(\mathbf{r}^\alpha)=\mathbf{V} + \bm{\Omega}\times\mathbf{r}^\alpha  \label{constraints4}
\end{equation}
for any position $\mathbf{r}^\alpha$ on the surface of the flagellum, and
\begin{equation}
\mathbf{v}(\mathbf{r}^\alpha)=\mathbf{V} + \bm{\Omega}\times\mathbf{r}^\alpha + \bm{\Omega}^{rel}\times\mathbf{r}^\alpha   \label{constraints5}
\end{equation}
for any position $\mathbf{r}^\alpha$ on the surface of the cell body.

%
%


Furthermore, the swimmer is force and torque free,
\begin{eqnarray}
0 &=& \sum_{\alpha=1}^{N_s}\mathbf{f}^{\alpha}, \label{constraints1} \\ 
0 &=& \sum_{\alpha=1}^{N_s}\mathbf{r}^{\alpha} \times \mathbf{f}^{\,\alpha} \label{constraints2}.
\end{eqnarray}

To find the unknown translational and rotational swimmer velocities, as well as the forces $\mathbf{f}^\alpha$ that determine the entire velocity field (Eq. ~\ref{velocity}), we follow a procedure similar to that we described previously in Ref. \cite{Martindale2016}, with a modification to incorporate static porous medium Stokeslets.
The kinematic boundary conditions (Eq. ~\ref{constraints3} - ~\ref{constraints5}) express the velocity at each Stokeslet position as a linear functions of $\mathbf{V}$ and $\bm{\Omega}$.  Therefore we can express them in matrix form as
\begin{equation}
\left[ \mathbf{v} \right] = \mathbf{K} \left[ \begin{array}{c} \mathbf{V} \\ \bm{\Omega} \end{array} \right] + \dot{\mathbf{L} }, \label{BCmatrix}
\end{equation}
where $\left[ \mathbf{v} \right] $ is a $3N$-element vector holding the components of the velocities at the $N$ positions $\mathbf{r}^\alpha$, and $\left[ \begin{array}{c} \mathbf{V}\\\bm{\Omega} \end{array} \right]$ is a 6-element vector holding the components of $\mathbf{V}$ and $\bm{\Omega}$.  
$\mathbf{K}$ is therefore a $3N\times6$  matrix, and explicitly, the row corresponding to the $i$th component of the $\alpha$th Stokeslet position on the cell body or flagellum is
\begin{equation}
\left[\delta_{i1} \,\,\,\,\,\, \delta_{i2} \,\,\,\,\,\, \delta_{i3} \,\,\,\,\,\, -\epsilon_{ij1} r^\alpha_j \,\,\,\,\,\, -\epsilon_{ij2} r^\alpha_j \,\,\,\,\,\, -\epsilon_{ij3} r^\alpha_j  \right],
\end{equation}
where $\delta_{ij}$ is the Kronecker delta and $\epsilon_{ijk}$ is the antisymmetric Levi-Civita tensor. The rows of $\mathbf{K}$ corresponding to Stokeslet positions in the porous medium are zero; this modification from \cite{Martindale2016} is what is needed to treat the static regularized Stokeslets in our problem.  
$\dot{\mathbf{L}}$ is a $3N$-element vector encoding the head rotation due to the motor, and explicitly, the element corresponding to the $i$th component of velocity at the $\alpha$th Stokeslet position on the cell body is
\begin{equation}
\epsilon_{ijk} \Omega^{rel}_{j} r^\alpha_k .
\end{equation}
The elements of $\dot{\mathbf{L}}$ corresponding to any Stokeslet position on the flagellum or in the porous medium are zero. 

The velocities $\left[ \mathbf{v} \right] $ can also be written as a linear function of the forces $\mathbf{f}^\alpha$ using Eq. ~\ref{velocity}, or in matrix form
\begin{equation}
\left[ \mathbf{v} \right] = \mathbf{G} \left[ \mathbf{f} \right], \label{velocitymatrix}
\end{equation} 
where $\left[ \mathbf{f} \right]$ is a $3N$-element vector consisting of the components of the forces $\mathbf{f}^\alpha$, and $\mathbf{G}$ is a $3 N \times 3N$ matrix.  Explicitly, the element of $\mathbf{G}$ giving the contribution to the $i$th component of the velocity at $\mathbf{r}^\alpha$ arising from the $j$th component of the $\beta$th regularized Stokeslet is $S_{ij}(\mathbf{r}^\alpha - \mathbf{r}^\beta, \epsilon^\beta)$. 

Combining Eqs. ~\ref{BCmatrix} and ~\ref{velocitymatrix}, one can solve for the forces as 
\begin{equation}
\left[ \mathbf{f} \right] = \mathbf{G}^{-1} \mathbf{K} \left[ \begin{array}{c} \mathbf{V} \\ \bm{\Omega} \end{array} \right] + \mathbf{G}^{-1} \dot{\mathbf{L} } \label{forcematrix}.
\end{equation}
The force and torque conditions (Eqs. ~\ref{constraints1} and ~\ref{constraints2} ) can be written as $0 = \mathbf{K}^\mathsf{T} \left[ \mathbf{f} \right]$, so 
\begin{equation}
0 = \mathbf{K}^\mathsf{T} \mathbf{G}^{-1} \mathbf{K} \left[ \begin{array}{c} \mathbf{V} \\ \bm{\Omega} \end{array} \right] + \mathbf{K}^\mathsf{T} \mathbf{G}^{-1} \dot{\mathbf{L} }.
\end{equation}
In this last equation, $ \mathbf{K}^\mathsf{T} \mathbf{G}^{-1} \mathbf{K}$ is a $6\times6$ matrix which can be inverted to solve for $\left[ \begin{array}{c} \mathbf{V} \\ \bm{\Omega} \end{array} \right]$.  Once solved for, $\left[ \begin{array}{c} \mathbf{V} \\ \bm{\Omega} \end{array} \right]$ determine the forces $\mathbf{f}^\alpha$ via Eq. ~\ref{forcematrix}, which then determine the entire velocity field via Eq. ~\ref{velocity}.
The average swimming speed of the swimmer, $\overline{\mathbf{V}}$ is then $(\mathbf{V}\cdot \hat{\bm{\Omega}})\hat{\bm{\Omega}}$ \cite{Meshkati2014}.

The most computationally expensive parts of the above method are the calculation of $\mathbf{G}^{-1} \mathbf{K}$ and $\mathbf{G}^{-1} \dot{\mathbf{L} }$, which amounts to 7 solutions of the linear equation   $\mathbf{G} \mathbf{x} = \mathbf{b}$ for $\mathbf{x}$ with the dense $3N\times 3N$ matrix $G$.  (Six solutions come from the columns of $\mathbf{K}$, and one from $\dot{\mathbf{L}}$.)  $N$ is typically very large (a few hundreds of thousands to millions) due to the static porous medium Stokeslets
and thus the standard schemes to solve for such large system of equations such as LU decomposition or GMRES become computationally very expensive. We use the GMRES algorithm, which involves computing the Krylov sub-spaces $\mathrm{Span}\lbrace \mathbf{b}, \mathbf{G}\mathbf{b}, \mathbf{G}^2\mathbf{b}, ... \rbrace$ created by successively multiplying $b$ with increasing powers of $\mathbf{G}$, and searching for an approximate solution $\hat{\mathbf{x}}$ in these sub-spaces such that the residual norm $|\mathbf{G}\hat{\mathbf{x}} - b|$ is within a prescribed tolerance \cite{Saad1986}.  To make our calculation numerically feasible, we implement a parallelized KIFMM as described in \cite{Rostami2016, Rostami2019} to find the matrix product $\mathbf{G}^{n} \mathbf{b}$ while avoiding explicit construction of $\mathbf{G}^{n}$.

We used 6759 regularized Stokeslets to describe the geometry of the bacterium. Of them, 4428 are on the flagellum with blob size $0.00405 \,\mu m$ and the rest on the swimmer body with blob size $0.01215 \, \mu m$. The swimmer body is modeled as a sphere of radius $0.5\,\mu m$ and the flagellar bundle (referred to simply as ``flagellum" throughout the paper) is modeled as a $0.7\, \mu m$ thick tapered helix of pitch $p=1.58\, \mu m$, diameter $d= 0.28\,\mu m$ ,and length $L = 2.97 \,\mu m$ \cite{Constantino2016}. The flagellum centerline ($\mathbf{r}_c(s)$) is described by 
\begin{equation}
\mathbf{r}_c(s) =
-s\hat{\mathbf{x}} +d/2\left(1-\exp^{-s^2/k^2_E}\right)\left[\cos(2\pi s/p)\hat{\mathbf{y}} + \sin(2\pi s/p)\hat{\mathbf{z}}\right], 
\end{equation}
where $s$ is the $x$ coordinate of a point on the centerline, and $k_E = p/2\pi$ is the characteristic length of the tapering region \cite{Martindale2016}.  

We reduce the number of regularized Stokeslets needed to represent the porous medium by only placing the random spatial distribution of Stokeslets within a layer along the boundary with the fluid pocket.  Since the flows are generated by the swimmer within the pocket, the fluid velocity quickly decays inside the porous medium.  

\begin{figure}
	\includegraphics[width=0.4\textwidth]{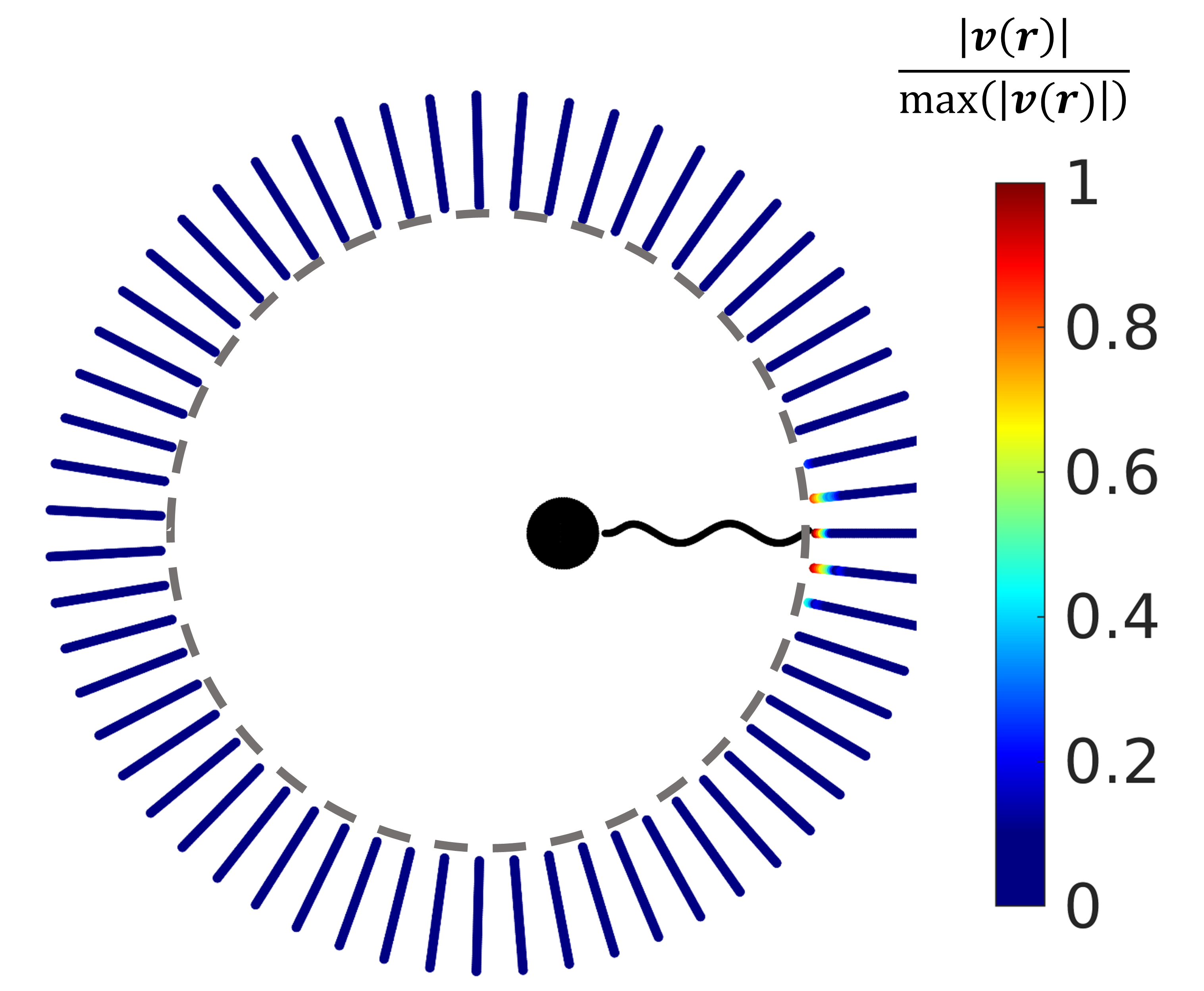} 
	\caption{Velocity field decay in the porous medium outside of the spherical fluid pocket (dotted line) within a $9\alpha^{-1}$ thick layer of regularized Stokeslets (shown to scale along radial sections). $\mathbf{v}$ is the flow velocity at position $\mathbf{r}$ in the porous medium, and $\mathrm{max}(|\mathbf{v}(r)|)$ is the largest magnitude of flow velocity on the pocket surface.  It can be seen that the velocity is largest near the flagellum, but decays rapidly outside the pocket. }
	\label{decay}
\end{figure}

\begin{figure}
	\includegraphics[width=0.6\columnwidth]{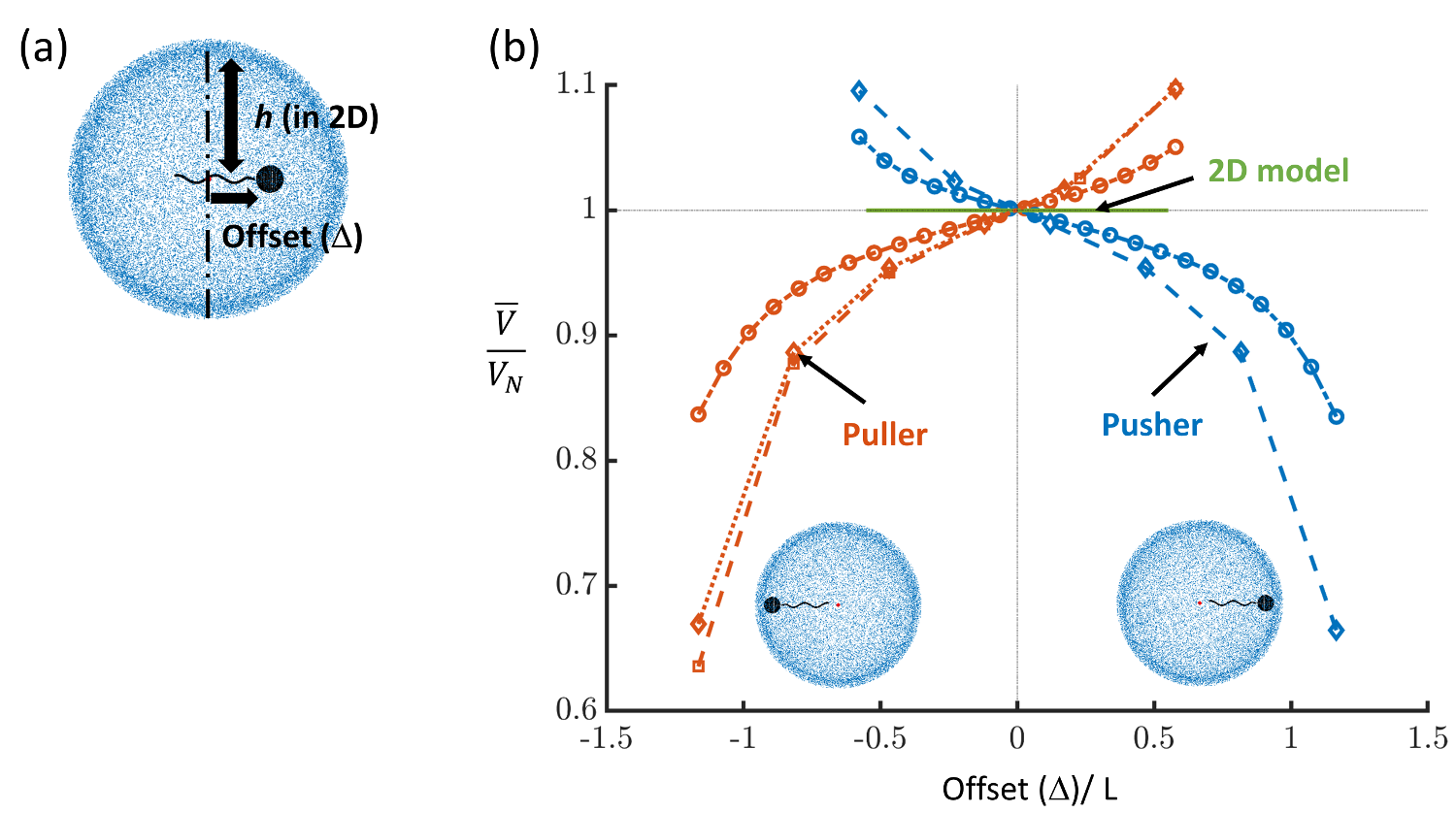} 
	\caption{(a) A swimmer inside spherical confinement with offset $\Delta$ in the swimming direction. Blue points are a shell of regularized Stokeslets representing gel surrounding a fluid pocket. (b) Velocity of  pusher and puller bacteria swimming inside spherical confinements of thickness ($\alpha^{-1}$, 3$\alpha^{-1}$, and 9$\alpha^{-1}$ represented by $\circ$, $\Diamond$, and $\square$ respectively, where $\alpha$ is the resistance of the corresponding Brinkman medium), compared to 2D confinement. We used 12076, 38414, and 142706 regularized Stokeslets to model the $\alpha^{-1}$, 3$\alpha^{-1}$, and 9$\alpha^{-1}$ thick porous shells. }
	\label{sphereconfineconv}
\end{figure}

To understand the layer thickness required to accurately represent the porous medium, we first found a layer thickness such that the velocity magnitude outside the layer is  $<90$\% of the largest magnitude flows at the inside surface of a spherical pocket (Fig. \ref{decay}).  This required a thickness of $1.575\, \mu m$, which is nine times the screening thickness $\alpha$ \cite{Durlofsky1987} of a Brinkman medium with matching resistive properties to the porous medium represented by our random distribution of regularized Stokeslets with density $\rho = 240 /{\mu m}^3$ and blob parameter $\epsilon = 0.01 \mu m$ \cite{Kamarapu2021}. 
In calculations of the swimming velocity and velocity field, we gradually increased the thickness of the shell from $\alpha^{-1}$, to $3 \alpha^{-1}$, to $9 \alpha^{-1}$ as shown in  Fig. \ref{sphereconfineconv}.  For Fig. \ref{sphereconfineconv}, at a thickness of $9 \alpha^{-1}$ the swimmer velocity changes by less than $\sim$ 5\% relative to a thickness of $3\alpha^{-1}$.  The converged results at thickness of $9 \alpha^{-1}$ are shown in Fig. 3 of the main text, and required 142,706 regularized Stokeslets to model the confinement. 
Note that the pusher solution for a thickness of $9 \alpha^{-1}$ is obtained from kinematic reversibility of the puller solution by reflecting it about $\Delta=0$.  However, the pusher solutions for thicknesses $\alpha^{-1}$ and $3 \alpha^{-1}$ were directly computed; they obey the same reflection property and confirm the validity of using kinematic reversibility to relate the pusher and puller solutions.

\section{Solution of Advection-Diffusion problem}
\label{A-D}

We first nondimensionalize the steady advection-diffusion equation : 
\begin{equation}\label{steadyAD}
\left(\textbf{v} - \textbf{V} - \bm{\Omega}\times r\right)\cdot \nabla C - D\nabla^2 C = 0
\end{equation} 
using length scales of the cell body radius $a$, translational and rotational velocity scales of the swimmer average speed $\overline{V}$ and rotational rate $\Omega$, respectively, and concentration scales of the ammonia concentration on the swimmer body surface.  In non-dimensional form, Eq. ~\ref{steadyAD} becomes
\begin{equation}\label{steadyND}
\left[Pe_U\left(\textbf{v}^* - \textbf{V}^*\right) - Pe_\Omega(\hat{\bm{\Omega}}\times \textbf{r}^*)\right]\cdot \nabla^* C^* = \mathbf{\nabla}^{*2}C^*, 
\end{equation}
where nondimensional quantities are indicated by asterisks, and the translational Peclet number $Pe_U$ and rotational Peclet number $Pe_\Omega$ are 
\begin{eqnarray}
Pe_U &=& \frac{a\overline{V}}{D}  \text{ and}\\
Pe_\Omega &=& \frac{a^2\Omega}{D} \text{ respectively}.
\end{eqnarray}

We prescribe a no flux condition on the surface of the flagellum and zero ammonia concentration on the far boundary of the advection-diffusion computational domain. Since we work in the frame of the flagellum, as the distance $\mathbf{r}$ from the swimmer increases, the rotational part of Eq. ~\ref{steadyAD} can become very large. We found that convergence of the numerical solution to this equation required a fine 3D computational mesh around the bacterium.  

Convergence was improved by decomposing the velocity and concentration fields into the velocity field $\mathbf{V}_s$ and concentration field $C_s$ of a sphere of radius $a$ at the same position and translating at the same mean velocity as the cell body ($\overline{\mathbf{V}}$), plus perturbations to these spherical solutions,  
\begin{eqnarray}\label{split}
\textbf{v} &=& \textbf{V}_s + \delta\textbf{v} \\
C &=& C_s + \delta C
\end{eqnarray}
The velocity field of a translating sphere $\textbf{V}_s$ in Stokes flow is well-known \cite{Happel1983}. The translating sphere concentration $C_s$ obeys
\begin{equation}\label{TSAD}
\left(\textbf{V}_s - \overline{\mathbf{V}}\right).\nabla C_s = D\nabla^2C_s 
\end{equation}
which can be efficiently solved in COMSOL. Note that the translating sphere concentration is
rotationally invariant about the axis of translation, \textit{i.e.},
\begin{equation}\label{rotinv}
\left(\bm{\Omega}\times\textbf{r}\right)\cdot\nabla C_s = 0.
\end{equation} 
Substituting Eq. ~\ref{split} into Eq. ~\ref{steadyAD} and using Eq.~\ref{rotinv}, we get an equation for the concentration perturbation $\delta C$ in non-dimensional form with reduced influence of the  rotational component:
\begin{equation}\label{finalAD}
\left[Pe_U\left(\textbf{v}^* - \textbf{V}^*\right) - Pe_\Omega\left(\hat{\bm{\Omega}}\times \textbf{r}^*\right) \right]\cdot\nabla^* \left(\delta C^*\right) 
= \nabla^{*2}\left(\delta C^*\right) + Pe_U\left[\left(\textbf{V}_s^* - \hat{\overline{\mathbf{V}}} \right)- \left(\textbf{v}^* - \textbf{V}^*\right)\right]\cdot\nabla^*C^*_s,
\end{equation}
which is an advection-diffusion equation for $\delta C^*$, with an additional source term (the second term on the right hand side) which can be computed from the translating sphere problem. 
We gradually refine the computational grid around the swimmer and numerically solve Eq.~\ref{finalAD} in COMSOL to obtain the converged fluid pocket geometry. We stop further refinement of the grid when the change in the fluid pocket volume is less than 5\% of the previous solution (the procedure to compute the volume difference is detailed in the next section).  
We find that solving Eq.~\ref{finalAD} numerically in COMSOL for $Pe_U = 6$ results in a converged pocket geometry using $\approx12 \times 10^6$ mesh elements, while the solution from the unmodified equation (Eq.~\ref{steadyND}) is still in a rapidly converging regime (pocket volume changing $\sim$ 70\% from the previous solution) at the same grid refinement. 

\section{Iterative self-consistent solution} 
\label{Iterative}

In this section, we provide details of the implementation  of the iterative procedure described in the main text to obtain a self-consistent solution of fluid pocket geometry.  Starting from the translational Peclet number, $Pe_U$, and the gait of the swimmer, $\bm{\Omega}_{rel}$, we first compute the unconfined swimming flows, and then solve the steady advection diffusion equation (Eq. ~\ref{finalAD},~\ref{TSAD}) { numerically in COMSOL to obtain the ammonia concentration field $C^*$ around the bacterium. We treat the region around the bacterium with $C^*>0.01$, corresponding to pH $>$ 4, as the fluid pocket.  
	Outside the fluid pocket is the mucus gel that confines the flow field, represented by a layer of randomly distributed static regularized Stokeslets \cite{Kamarapu2021} in a shell.

	Then we follow the procedure described in Sec. \ref{HDmodel} 
	to compute the flow field inside the fluid pocket.  This flow field is then used to re-solve the advection-diffusion problem, which gives a new pocket geometry.  We repeat the above steps until convergence.
	A converged self-consistent solution of fluid pocket geometry confines the flow field around the bacterium in such a way that it results in the same fluid pocket after solving the advection-diffusion problem.  Our metric to check the convergence is the volumetric difference between fluid pockets at successive iterations, where the volumetric difference is unsigned, i.e., it is the volume of the space between the boundary of one pocket and the other pocket, counting all such volume as positive contributions to the change.
	To compute this metric, we randomly scatter points in the domain at a sufficiently high density to fill in the intricacies of the fluid pocket, and then count the number of points that are in the volume between the two pocket boundaries.  We normalize that volume using the volume of the pocket of the previous iteration, also measured by counting randomly scattered points. 
	We stop the iterative procedure when the volume difference metric is under 5$\%$.  Starting from the second iteration, we prescribe the fluid velocity to be zero outside the porous medium shell while solving the advection diffusion equation. 
	As the number of Stokeslets describing the gel can be large, we use a $\alpha^{-1}$ thick shell in the initial few iterations to save computational time.  After that, during the hydrodnyamic calculations we tested shells of increasing thickness ($3 \alpha^{-1}$, $9 \alpha^{-1}$) until the swimming velocity changed by less than 5\% relative to the previous thickness.

	For Peclet number $Pe_U = 6$ (roughly 4000 times the bacterial Peclet number) the fluid flow field is very sensitive to the fluid pocket geometry as its boundary is very close to the swimmer. Hence, convergence to the self-consistent solution takes many iterations.  The converged porous medium shell had a thickness of $9 \alpha^{-1}$, which required 8 million regularized Stokeslets to represent the porous medium.

	For low and medium $Pe$ numbers, fewer iterations are required since the advective effects are reduced. Convergence at low $Pe = 1.52 \times 10^{-3}$ is almost immediate within two iterations. Owing to the large fluid pocket sizes in these regimes, the self-consistent swimming velocity changed by less than 1$\%$ when the mucus shell thickness is increased from $1\alpha^{-1}$ to $3\alpha^{-1}$, so the final reported results use a shell thickness of $3 \alpha^{-1}$.  The number of porous medium stokeslets required to model the mucus gel confinement at low and medium $Pe$ numbers are 13 and 5 million, respectively.
	
\section{Spherical pocket with offset spherical squirmer}
\label{Bispherical}

In this section we describe the analytic solution of Stokes flow and swimming speed of a spherical squirmer with tangential prescribed slip velocities on its surface, swimming in spherical confinement.  Unlike in previous studies \cite{Reigh2017, Nganguia2020}, the squirmer can take an off-center position in the spherical pocket, which allows us to study the effect of fore-aft clearance.

The solution method takes advantage of the bispherical coordinate system $(\xi, \eta, \phi)$, in which surfaces of constant $\xi$ are non-intersecting spheres of differing sizes at different locations along the $z$-axis (See Fig. 4 of main text). The solution is a relatively straightforward extension of the solution presented in \cite{Jabbarzadeh2018} for a spherical squirmer approaching a solid sphere.

When the squirmer and pocket are both centered along the $z$-axis, the problem and solution are axisymmetric (independent of azimuthal angle $\phi$).  In that case, the solution to Stokes equations (Eq. 1 of the main text) can be expressed in terms of a streamfunction $\psi$ as $\mathbf{v} = \mathrm{curl}(\psi \mathbf{e}_\phi)$, with \cite{Stimson1926}
\begin{subeqnarray}
	\psi &=& \big(\cosh \xi -\mu \big)^{-3/2}X ,\\
	X &=&\sum_{n=1}^{\infty} U_n(\xi)V_n(\mu), \label{Psi} \\
	U_n(\xi)&=&\mathcal{A}_n\cosh \left((n-1/2)\xi\right)+\mathcal{B}_n\sinh\left((n-1/2)\xi\right)  \\ \nonumber
	&+&\mathcal{C}_n\cosh\left((n+3/2)\xi\right)+\mathcal{D}_n\sinh \left((n+3/2)\xi\right), \label{Un}
\end{subeqnarray}
where $V_n(\mu) = P_{n-1}(\mu) - P_{n+1}(\mu)$ in terms of the Legendre functions $P_n(\mu)$, $\mu = \cos \eta$, and $\mathcal{A}_n,\mathcal{B}_n,\mathcal{C}_n$, $\mathcal{D}_n$ are constants to be determined by imposing boundary conditions.

If a squirmer of radius $r_s$ centered at position $z_s$ along the z-axis is inside a pocket of radius $r_p$ centered at position $z_p$, then the clearance $c$ (Fig. 4 of main text) between the squirmer and pocket surfaces is $c= (z_s - r_s) - (z_p- r_p)$.  Up to a scaling factor $C$, the situation can be described using two surfaces of constant $\xi$, $\xi = \alpha >0$ for the squirmer and $\xi=\beta >0$ for the pocket.  Then $r_s = C \, \mathrm{csch} \, \alpha$, $z_s = C \coth \alpha$, $r_p = C \, \mathrm{csch} \, \beta$, and $z_p = C \coth \beta$. These relations can be inverted to find $\alpha$, $\beta$, and $C$ in terms of $r_s$, $r_p$, and $c$:
\begin{eqnarray}
\alpha &=& \cosh^{-1}(\frac{c}{r_s}\frac{2(r_p/r_s-1)-c/r_s}{2(r_p/r_s-1-c/r_s)}),\\
\beta  &=& \cosh^{-1}(1-\frac{2(r_p/r_s-1-c/r_s)-(c/r_s)^2}{2(r_p/r_s)(r_p/r_s-1-c/r_s)})
\\
\frac{C}{r_s} &=& \sqrt{(\frac{(c/r_s)(c/r_s-2r_p/r_s+2)}{2(c/r_s-r_p/r_s+1)})^2-1}	\label{parameter}
\end{eqnarray}

The pocket surface is stationary and the no-slip boundary conditions yield
\begin{eqnarray}
\psi(\xi=\beta) &=& 0\\
\partial_\xi \psi(\xi=\beta) &=& 0.
\end{eqnarray}
The squirmer has a prescribed tangential slip velocity at its surface, parameterized by $B_1$ and $B_2$, and an overall translational (swimming) velocity $U$ along the z axis.  The swimming velocity of an unconfined squirmer is $U_N =2 B_1/3 $, while the ratio $B_2/B_1$ determines whether the squirmer is a pusher or a puller.  For $B_2/B_1 < 0$, the squirmer is a pusher, while for $B_2/B_1 > 0$, the squirmer is a puller. In a spherical coordinate system centered at the squirmer, with polar angle defined from the positive z-axis, the boundary condition at the squirmer surface is 
\begin{subeqnarray}
	u_r &=& U\cos \theta,\\
	u_\theta &=& (-U+B_1) \sin \theta + B_2 \sin(2\theta)/2. \label{GBC}
\end{subeqnarray}
In the bispherical coordinate system, these boundary conditions are equivalent to
\begin{subeqnarray}
	\partial_\xi \psi(\xi=\alpha, \eta) &=& - u_\theta \frac{\rho z r_s}{C}\\
	\partial_\eta \psi(\xi=\alpha, \eta) &=& - u_r \frac{\rho z r_s}{C}, \label{squirmerBC}
\end{subeqnarray}

Together, these boundary conditions can be used to solve for the coefficients $\mathcal{A}_n,\mathcal{B}_n,\mathcal{C}_n$ for each $n$.  These satisfy the linear set of equations
\begin{subeqnarray}
	\mathcal{A}_n \cosh(n-\frac{1}{2})\alpha &+& \mathcal{B}_n \sinh(n-\frac{1}{2})\alpha + \mathcal{C}_n \cosh(n+\frac{3}{2})\alpha + \mathcal{D}_n \sinh(n+\frac{3}{2})\alpha  \nonumber \\
	&=&-k U\Big[(2n+3)e^{-(n-\frac{1}{2})\alpha} - (2n-1)e^{-(n+\frac{3}{2})\alpha}\Big] \label{swimmer1} \\ 
	(2n-1)[\mathcal{A}_n \sinh (n -\frac{1}{2})\alpha &+& \mathcal{B}_n \cosh(n-\frac{1}{2})\alpha]
	+ (2n+3)[\mathcal{C}_n \sinh(n+\frac{3}{2})\alpha + \mathcal{D}_n \cosh(n+\frac{3}{2})\alpha ] \nonumber\\
	&=& U H_n(\alpha) + G_n(\alpha)  \\ 
	\mathcal{A}_n \cosh(n-\frac{1}{2})\beta &+& \mathcal{B}_n \sinh(n-\frac{1}{2})\beta + \mathcal{C}_n \cosh(n+\frac{3}{2})\beta + \mathcal{D}_n \sinh(n+\frac{3}{2})\beta  \nonumber \\
	&=& -k V_t\Big[(2n+3)e^{-(n-\frac{1}{2})\beta} - (2n-1)e^{-(n+\frac{3}{2})\beta}\Big]  \\ 
	(2n-1)\Big[\mathcal{A}_n \sinh(n-\frac{1}{2})\beta &+& \mathcal{B}_n \cosh(n-\frac{1}{2})\beta\Big] + (2n+3)\Big[\mathcal{C}_n \sinh(n+\frac{3}{2})\beta+ \mathcal{D}_n \cosh(n+\frac{3}{2})\beta\Big] \nonumber\\
	&=& -kV_t(2n-1)(2n+3)\Big[e^{-(n-\frac{1}{2})\beta} - e^{-(n+\frac{3}{2})\beta}\Big], \label{Expansion_localized}
\end{subeqnarray}
where
\begin{equation}
k=\frac{c^2 n(n+1)}{\sqrt{2}(2n-1)(2n+1)(2n+3)}.
\end{equation}
In these formulas,  $H_n(\alpha)$ and $G_n(\alpha)$ are set by the values of $B_1$ and $B_2$, and are determined by a Taylor-Legendre expansion of Eqs. \ref{squirmerBC}, as described in \cite{Jabbarzadeh2018}. The details are not illuminating so we do not repeat them here, but note that we use exactly the same notation here and in that paper, with the exception that (capital) $C$ in this paper is the equivalent of (lowercase) $c$ in \cite{Jabbarzadeh2018}. 

Given $r_s$, $r_p$, and $c$, we find the values of $\alpha$, $\beta$, and $C$ using Eqs. \ref{parameter}.  Then, for  specific values of $B_1$ and $B_2$ we solve Eqs. \ref{Expansion_localized} for the coefficients $\mathcal{A}_n,\mathcal{B}_n,\mathcal{C}_n$, and $\mathcal{D}_n$. Once these coefficients are known, we can calculate the force on the squirmer as 
\begin{equation}
F_s = -\kappa \frac{2\pi\sqrt{2}}{c}\sum_{n=1}^{\infty} (2n+1)\big(\mathcal{A}_n+\mathcal{B}_n+\mathcal{C}_n+\mathcal{D}_n\big),
\label{force}
\end{equation}
where $\kappa$ is the fluid viscosity.
Typically the magnitude of the coefficients decay exponentially with $n$, and we truncate the solution when the last term in the series Eq. \ref{force} is less than $10^{-15}$ of the total force. 
The expression for the force yields a linear relation $F_s = R_1 U + R_2$, where the $R_i$ are constants.  Finally we solve for the confined swimming velocity $U$ by imposing the force-free condition on the squirmer, $F_s=0$.

The data in Fig. 4c of the main text were calculated for $r_s = 1, r_p = 10, B_2/B_1 = -1,-2,-3,-5,-10$.  Note that in this Appendix we used $U$ for the swimming speed to match the notation of \cite{Jabbarzadeh2018}.  In the main text we use $V$ (equal to $U$ in this Appendix) for the swimming speed and $V_N$ (equal to $U_N$ in this Appendix) for the unconfined swimming speed.


\begin{thebibliography}{96}%
	\makeatletter
	\providecommand \@ifxundefined [1]{%
		\@ifx{#1\undefined}
	}%
	\providecommand \@ifnum [1]{%
		\ifnum #1\expandafter \@firstoftwo
		\else \expandafter \@secondoftwo
		\fi
	}%
	\providecommand \@ifx [1]{%
		\ifx #1\expandafter \@firstoftwo
		\else \expandafter \@secondoftwo
		\fi
	}%
	\providecommand \natexlab [1]{#1}%
	\providecommand \enquote  [1]{``#1''}%
	\providecommand \bibnamefont  [1]{#1}%
	\providecommand \bibfnamefont [1]{#1}%
	\providecommand \citenamefont [1]{#1}%
	\providecommand \href@noop [0]{\@secondoftwo}%
	\providecommand \href [0]{\begingroup \@sanitize@url \@href}%
	\providecommand \@href[1]{\@@startlink{#1}\@@href}%
	\providecommand \@@href[1]{\endgroup#1\@@endlink}%
	\providecommand \@sanitize@url [0]{\catcode `\\12\catcode `\$12\catcode
		`\&12\catcode `\#12\catcode `\^12\catcode `\_12\catcode `\%12\relax}%
	\providecommand \@@startlink[1]{}%
	\providecommand \@@endlink[0]{}%
	\providecommand \url  [0]{\begingroup\@sanitize@url \@url }%
	\providecommand \@url [1]{\endgroup\@href {#1}{\urlprefix }}%
	\providecommand \urlprefix  [0]{URL }%
	\providecommand \Eprint [0]{\href }%
	\providecommand \doibase [0]{https://doi.org/}%
	\providecommand \selectlanguage [0]{\@gobble}%
	\providecommand \bibinfo  [0]{\@secondoftwo}%
	\providecommand \bibfield  [0]{\@secondoftwo}%
	\providecommand \translation [1]{[#1]}%
	\providecommand \BibitemOpen [0]{}%
	\providecommand \bibitemStop [0]{}%
	\providecommand \bibitemNoStop [0]{.\EOS\space}%
	\providecommand \EOS [0]{\spacefactor3000\relax}%
	\providecommand \BibitemShut  [1]{\csname bibitem#1\endcsname}%
	\let\auto@bib@innerbib\@empty
	\bibitem [{\citenamefont {Bechinger}\ \emph {et~al.}(2016)\citenamefont
		{Bechinger}, \citenamefont {{Di Leonardo}}, \citenamefont {L{\"{o}}wen},
		\citenamefont {Reichhardt}, \citenamefont {Volpe},\ and\ \citenamefont
		{Volpe}}]{Bechinger2016}%
	\BibitemOpen
	\bibfield  {author} {\bibinfo {author} {\bibfnamefont {C.}~\bibnamefont
			{Bechinger}}, \bibinfo {author} {\bibfnamefont {R.}~\bibnamefont {{Di
					Leonardo}}}, \bibinfo {author} {\bibfnamefont {H.}~\bibnamefont
			{L{\"{o}}wen}}, \bibinfo {author} {\bibfnamefont {C.}~\bibnamefont
			{Reichhardt}}, \bibinfo {author} {\bibfnamefont {G.}~\bibnamefont {Volpe}},\
		and\ \bibinfo {author} {\bibfnamefont {G.}~\bibnamefont {Volpe}},\ }\bibfield
	{title} {\bibinfo {title} {{Active particles in complex and crowded
				environments}},\ }\href {https://doi.org/10.1103/RevModPhys.88.045006}
	{\bibfield  {journal} {\bibinfo  {journal} {Reviews of Modern Physics}\
		}\textbf {\bibinfo {volume} {88}},\ \bibinfo {pages} {045006} (\bibinfo
		{year} {2016})}\BibitemShut {NoStop}%
	\bibitem [{\citenamefont {Lauga}\ and\ \citenamefont
		{Powers}(2009)}]{Lauga2009}%
	\BibitemOpen
	\bibfield  {author} {\bibinfo {author} {\bibfnamefont {E.}~\bibnamefont
			{Lauga}}\ and\ \bibinfo {author} {\bibfnamefont {T.~R.}\ \bibnamefont
			{Powers}},\ }\bibfield  {title} {\bibinfo {title} {{The hydrodynamics of
				swimming microorganisms}},\ }\href
	{https://iopscience.iop.org/article/10.1088/0034-4885/72/9/096601
		https://iopscience.iop.org/article/10.1088/0034-4885/72/9/096601/meta}
	{\bibfield  {journal} {\bibinfo  {journal} {Reports on Progress in Physics}\
		}\textbf {\bibinfo {volume} {72}},\ \bibinfo {pages} {096601} (\bibinfo
		{year} {2009})}\BibitemShut {NoStop}%
	\bibitem [{\citenamefont {Spagnolie}\ and\ \citenamefont
		{Lauga}(2012)}]{Spagnolie2012}%
	\BibitemOpen
	\bibfield  {author} {\bibinfo {author} {\bibfnamefont {S.~E.}\ \bibnamefont
			{Spagnolie}}\ and\ \bibinfo {author} {\bibfnamefont {E.}~\bibnamefont
			{Lauga}},\ }\bibfield  {title} {\bibinfo {title} {{Hydrodynamics of
				self-propulsion near a boundary: predictions and accuracy of far-field
				approximations}},\ }\href {https://doi.org/10.1017/jfm.2012.101} {\bibfield
		{journal} {\bibinfo  {journal} {Journal of Fluid Mechanics}\ }\textbf
		{\bibinfo {volume} {700}},\ \bibinfo {pages} {105} (\bibinfo {year}
		{2012})}\BibitemShut {NoStop}%
	\bibitem [{\citenamefont {Elgeti}\ \emph {et~al.}(2015)\citenamefont {Elgeti},
		\citenamefont {Winkler},\ and\ \citenamefont {Gompper}}]{Elgeti2015}%
	\BibitemOpen
	\bibfield  {author} {\bibinfo {author} {\bibfnamefont {J.}~\bibnamefont
			{Elgeti}}, \bibinfo {author} {\bibfnamefont {R.~G.}\ \bibnamefont
			{Winkler}},\ and\ \bibinfo {author} {\bibfnamefont {G.}~\bibnamefont
			{Gompper}},\ }\bibfield  {title} {\bibinfo {title} {{Physics of microswimmers
				- Single particle motion and collective behavior: A review}},\ }\href
	{https://doi.org/10.1088/0034-4885/78/5/056601} {\bibfield  {journal}
		{\bibinfo  {journal} {Reports on Progress in Physics}\ }\textbf {\bibinfo
			{volume} {78}},\ \bibinfo {pages} {056601} (\bibinfo {year}
		{2015})}\BibitemShut {NoStop}%
	\bibitem [{\citenamefont {Berg}\ and\ \citenamefont
		{Anderson}(1973)}]{Berg1973}%
	\BibitemOpen
	\bibfield  {author} {\bibinfo {author} {\bibfnamefont {H.~C.}\ \bibnamefont
			{Berg}}\ and\ \bibinfo {author} {\bibfnamefont {R.~A.}\ \bibnamefont
			{Anderson}},\ }\bibfield  {title} {\bibinfo {title} {{Bacteria swim by
				rotating their flagellar filaments}},\ }\href
	{https://doi.org/10.1038/245380a0} {\bibfield  {journal} {\bibinfo  {journal}
			{Nature}\ }\textbf {\bibinfo {volume} {245}},\ \bibinfo {pages} {380}
		(\bibinfo {year} {1973})}\BibitemShut {NoStop}%
	\bibitem [{\citenamefont {Walther}\ and\ \citenamefont
		{M{\"{u}}ller}(2013)}]{Walther2013}%
	\BibitemOpen
	\bibfield  {author} {\bibinfo {author} {\bibfnamefont {A.}~\bibnamefont
			{Walther}}\ and\ \bibinfo {author} {\bibfnamefont {A.~H.}\ \bibnamefont
			{M{\"{u}}ller}},\ }\bibfield  {title} {\bibinfo {title} {{Janus particles:
				Synthesis, self-assembly, physical properties, and applications}},\ }\href
	{https://doi.org/10.1021/cr300089t} {\bibfield  {journal} {\bibinfo
			{journal} {Chemical Reviews}\ }\textbf {\bibinfo {volume} {113}},\ \bibinfo
		{pages} {5194} (\bibinfo {year} {2013})}\BibitemShut {NoStop}%
	\bibitem [{\citenamefont {Sanchez}\ \emph {et~al.}(2012)\citenamefont
		{Sanchez}, \citenamefont {Chen}, \citenamefont {Decamp}, \citenamefont
		{Heymann},\ and\ \citenamefont {Dogic}}]{Sanchez2012}%
	\BibitemOpen
	\bibfield  {author} {\bibinfo {author} {\bibfnamefont {T.}~\bibnamefont
			{Sanchez}}, \bibinfo {author} {\bibfnamefont {D.~T.}\ \bibnamefont {Chen}},
		\bibinfo {author} {\bibfnamefont {S.~J.}\ \bibnamefont {Decamp}}, \bibinfo
		{author} {\bibfnamefont {M.}~\bibnamefont {Heymann}},\ and\ \bibinfo {author}
		{\bibfnamefont {Z.}~\bibnamefont {Dogic}},\ }\bibfield  {title} {\bibinfo
		{title} {{Spontaneous motion in hierarchically assembled active matter}},\
	}\href {https://doi.org/10.1038/nature11591} {\bibfield  {journal} {\bibinfo
			{journal} {Nature}\ }\textbf {\bibinfo {volume} {491}},\ \bibinfo {pages}
		{431} (\bibinfo {year} {2012})}\BibitemShut {NoStop}%
	\bibitem [{\citenamefont {Chen}\ \emph {et~al.}(2020)\citenamefont {Chen},
		\citenamefont {Xu}, \citenamefont {Zhu}, \citenamefont {Dai},\ and\
		\citenamefont {Yan}}]{Chen2020}%
	\BibitemOpen
	\bibfield  {author} {\bibinfo {author} {\bibfnamefont {P.}~\bibnamefont
			{Chen}}, \bibinfo {author} {\bibfnamefont {Z.}~\bibnamefont {Xu}}, \bibinfo
		{author} {\bibfnamefont {G.}~\bibnamefont {Zhu}}, \bibinfo {author}
		{\bibfnamefont {X.}~\bibnamefont {Dai}},\ and\ \bibinfo {author}
		{\bibfnamefont {L.~T.}\ \bibnamefont {Yan}},\ }\bibfield  {title} {\bibinfo
		{title} {{Cellular Uptake of Active Particles}},\ }\href
	{https://doi.org/10.1103/PhysRevLett.124.198102} {\bibfield  {journal}
		{\bibinfo  {journal} {Physical Review Letters}\ }\textbf {\bibinfo {volume}
			{124}},\ \bibinfo {pages} {198102} (\bibinfo {year} {2020})}\BibitemShut
	{NoStop}%
	\bibitem [{\citenamefont {Be'er}\ and\ \citenamefont {Ariel}(2019)}]{Beer2019}%
	\BibitemOpen
	\bibfield  {author} {\bibinfo {author} {\bibfnamefont {A.}~\bibnamefont
			{Be'er}}\ and\ \bibinfo {author} {\bibfnamefont {G.}~\bibnamefont {Ariel}},\
	}\bibfield  {title} {\bibinfo {title} {{A statistical physics view of
				swarming bacteria}},\ }\href@noop {} {\bibfield  {journal} {\bibinfo
			{journal} {Movement Ecology}\ }\textbf {\bibinfo {volume} {7}},\ \bibinfo
		{pages} {1} (\bibinfo {year} {2019})}\BibitemShut {NoStop}%
	\bibitem [{\citenamefont {Schwarz-Linek}\ \emph {et~al.}(2012)\citenamefont
		{Schwarz-Linek}, \citenamefont {Valeriani}, \citenamefont {Cacciuto},
		\citenamefont {Cates}, \citenamefont {Marenduzzo}, \citenamefont {Morozov},\
		and\ \citenamefont {Poon}}]{Schwarz-Linek2012}%
	\BibitemOpen
	\bibfield  {author} {\bibinfo {author} {\bibfnamefont {J.}~\bibnamefont
			{Schwarz-Linek}}, \bibinfo {author} {\bibfnamefont {C.}~\bibnamefont
			{Valeriani}}, \bibinfo {author} {\bibfnamefont {A.}~\bibnamefont {Cacciuto}},
		\bibinfo {author} {\bibfnamefont {M.~E.}\ \bibnamefont {Cates}}, \bibinfo
		{author} {\bibfnamefont {D.}~\bibnamefont {Marenduzzo}}, \bibinfo {author}
		{\bibfnamefont {A.~N.}\ \bibnamefont {Morozov}},\ and\ \bibinfo {author}
		{\bibfnamefont {W.~C.}\ \bibnamefont {Poon}},\ }\bibfield  {title} {\bibinfo
		{title} {{Phase separation and rotor self-assembly in active particle
				suspensions}},\ }\href {https://doi.org/10.1073/pnas.1116334109} {\bibfield
		{journal} {\bibinfo  {journal} {Proceedings of the National Academy of
				Sciences of the United States of America}\ }\textbf {\bibinfo {volume}
			{109}},\ \bibinfo {pages} {4052} (\bibinfo {year} {2012})}\BibitemShut
	{NoStop}%
	\bibitem [{\citenamefont {Michalec}\ \emph {et~al.}(2017)\citenamefont
		{Michalec}, \citenamefont {Fouxon}, \citenamefont {Souissi},\ and\
		\citenamefont {Holzner}}]{Michalec2017}%
	\BibitemOpen
	\bibfield  {author} {\bibinfo {author} {\bibfnamefont {F.~G.}\ \bibnamefont
			{Michalec}}, \bibinfo {author} {\bibfnamefont {I.}~\bibnamefont {Fouxon}},
		\bibinfo {author} {\bibfnamefont {S.}~\bibnamefont {Souissi}},\ and\ \bibinfo
		{author} {\bibfnamefont {M.}~\bibnamefont {Holzner}},\ }\bibfield  {title}
	{\bibinfo {title} {{Zooplankton can actively adjust their motility to
				turbulent flow}},\ }\href {https://doi.org/10.1073/pnas.1708888114}
	{\bibfield  {journal} {\bibinfo  {journal} {Proceedings of the National
				Academy of Sciences of the United States of America}\ }\textbf {\bibinfo
			{volume} {114}},\ \bibinfo {pages} {E11199} (\bibinfo {year}
		{2017})}\BibitemShut {NoStop}%
	\bibitem [{\citenamefont {Go{\'{z}}dziejewska}\ \emph
		{et~al.}(2019)\citenamefont {Go{\'{z}}dziejewska}, \citenamefont
		{Gwo{\'{z}}dzik}, \citenamefont {Kulesza}, \citenamefont {Bramowicz},\ and\
		\citenamefont {Kosza{\l}ka}}]{Gozdziejewska2019}%
	\BibitemOpen
	\bibfield  {author} {\bibinfo {author} {\bibfnamefont {A.~M.}\ \bibnamefont
			{Go{\'{z}}dziejewska}}, \bibinfo {author} {\bibfnamefont {M.}~\bibnamefont
			{Gwo{\'{z}}dzik}}, \bibinfo {author} {\bibfnamefont {S.}~\bibnamefont
			{Kulesza}}, \bibinfo {author} {\bibfnamefont {M.}~\bibnamefont {Bramowicz}},\
		and\ \bibinfo {author} {\bibfnamefont {J.}~\bibnamefont {Kosza{\l}ka}},\
	}\bibfield  {title} {\bibinfo {title} {{Effects of suspended micro- and
				nanoscale particles on zooplankton functional diversity of drainage system
				reservoirs at an open-pit mine}},\ }\href
	{https://doi.org/10.1038/s41598-019-52542-6} {\bibfield  {journal} {\bibinfo
			{journal} {Scientific Reports}\ }\textbf {\bibinfo {volume} {9}},\ \bibinfo
		{pages} {1} (\bibinfo {year} {2019})}\BibitemShut {NoStop}%
	\bibitem [{\citenamefont {Fenchel}(1986)}]{Fenchel1986}%
	\BibitemOpen
	\bibfield  {author} {\bibinfo {author} {\bibfnamefont {T.}~\bibnamefont
			{Fenchel}},\ }\bibfield  {title} {\bibinfo {title} {{The Ecology of
				Heterotrophic Microflagellates}},\ }in\ \href
	{https://doi.org/10.1007/978-1-4757-0611-6_2} {\emph {\bibinfo {booktitle}
			{Advances in Microbial Ecology}}},\ \bibinfo {editor} {edited by\ \bibinfo
		{editor} {\bibfnamefont {K.~C.}\ \bibnamefont {Marshall}}}\ (\bibinfo
	{publisher} {Springer, Boston, MA},\ \bibinfo {year} {1986})\ pp.\ \bibinfo
	{pages} {57--97}\BibitemShut {NoStop}%
	\bibitem [{\citenamefont {Ki{\o}rboe}(2016)}]{Kiorboe2016}%
	\BibitemOpen
	\bibfield  {author} {\bibinfo {author} {\bibfnamefont {T.}~\bibnamefont
			{Ki{\o}rboe}},\ }\bibfield  {title} {\bibinfo {title} {{Fluid dynamic
				constraints on resource acquisition in small pelagic organisms}},\
	}\href@noop {} {\bibfield  {journal} {\bibinfo  {journal} {The European
				Physical Journal Special Topics}\ }\textbf {\bibinfo {volume} {225}},\
		\bibinfo {pages} {669} (\bibinfo {year} {2016})}\BibitemShut {NoStop}%
	\bibitem [{\citenamefont {Azam}\ \emph {et~al.}(1994)\citenamefont {Azam},
		\citenamefont {Smith}, \citenamefont {Steward},\ and\ \citenamefont
		{Hagstr{\"{o}}m}}]{Azam1994}%
	\BibitemOpen
	\bibfield  {author} {\bibinfo {author} {\bibfnamefont {F.}~\bibnamefont
			{Azam}}, \bibinfo {author} {\bibfnamefont {D.~C.}\ \bibnamefont {Smith}},
		\bibinfo {author} {\bibfnamefont {G.~F.}\ \bibnamefont {Steward}},\ and\
		\bibinfo {author} {\bibfnamefont {{\AA}.}~\bibnamefont {Hagstr{\"{o}}m}},\
	}\bibfield  {title} {\bibinfo {title} {{Bacteria-organic matter coupling and
				its significance for oceanic carbon cycling}},\ }\href
	{https://doi.org/10.1007/BF00166806} {\bibfield  {journal} {\bibinfo
			{journal} {Microbial Ecology}\ }\textbf {\bibinfo {volume} {28}},\ \bibinfo
		{pages} {167} (\bibinfo {year} {1994})}\BibitemShut {NoStop}%
	\bibitem [{\citenamefont {Xu}\ \emph {et~al.}(2018)\citenamefont {Xu},
		\citenamefont {Hou}, \citenamefont {Wattanatorn}, \citenamefont {Wang},
		\citenamefont {Yang}, \citenamefont {Zhao}, \citenamefont {Yu}, \citenamefont
		{Tseng}, \citenamefont {Jonas},\ and\ \citenamefont {Weiss}}]{Xu2018}%
	\BibitemOpen
	\bibfield  {author} {\bibinfo {author} {\bibfnamefont {X.}~\bibnamefont
			{Xu}}, \bibinfo {author} {\bibfnamefont {S.}~\bibnamefont {Hou}}, \bibinfo
		{author} {\bibfnamefont {N.}~\bibnamefont {Wattanatorn}}, \bibinfo {author}
		{\bibfnamefont {F.}~\bibnamefont {Wang}}, \bibinfo {author} {\bibfnamefont
			{Q.}~\bibnamefont {Yang}}, \bibinfo {author} {\bibfnamefont {C.}~\bibnamefont
			{Zhao}}, \bibinfo {author} {\bibfnamefont {X.}~\bibnamefont {Yu}}, \bibinfo
		{author} {\bibfnamefont {H.~R.}\ \bibnamefont {Tseng}}, \bibinfo {author}
		{\bibfnamefont {S.~J.}\ \bibnamefont {Jonas}},\ and\ \bibinfo {author}
		{\bibfnamefont {P.~S.}\ \bibnamefont {Weiss}},\ }\bibfield  {title} {\bibinfo
		{title} {{Precision-Guided Nanospears for Targeted and High-Throughput
				Intracellular Gene Delivery}},\ }\href
	{https://doi.org/10.1021/acsnano.8b00763} {\bibfield  {journal} {\bibinfo
			{journal} {ACS Nano}\ }\textbf {\bibinfo {volume} {12}},\ \bibinfo {pages}
		{4503} (\bibinfo {year} {2018})}\BibitemShut {NoStop}%
	\bibitem [{\citenamefont {Stenhammar}\ \emph {et~al.}(2015)\citenamefont
		{Stenhammar}, \citenamefont {Wittkowski}, \citenamefont {Marenduzzo},\ and\
		\citenamefont {Cates}}]{Stenhammar2015}%
	\BibitemOpen
	\bibfield  {author} {\bibinfo {author} {\bibfnamefont {J.}~\bibnamefont
			{Stenhammar}}, \bibinfo {author} {\bibfnamefont {R.}~\bibnamefont
			{Wittkowski}}, \bibinfo {author} {\bibfnamefont {D.}~\bibnamefont
			{Marenduzzo}},\ and\ \bibinfo {author} {\bibfnamefont {M.~E.}\ \bibnamefont
			{Cates}},\ }\bibfield  {title} {\bibinfo {title} {{Activity-induced phase
				separation and self-assembly in mixtures of active and passive particles}},\
	}\href {https://journals.aps.org/prl/abstract/10.1103/PhysRevLett.114.018301}
	{\bibfield  {journal} {\bibinfo  {journal} {Physical Review Letters}\
		}\textbf {\bibinfo {volume} {114}},\ \bibinfo {pages} {018301} (\bibinfo
		{year} {2015})}\BibitemShut {NoStop}%
	\bibitem [{\citenamefont {Kim}\ \emph {et~al.}(2013)\citenamefont {Kim},
		\citenamefont {Laschi},\ and\ \citenamefont {Trimmer}}]{Kim2013}%
	\BibitemOpen
	\bibfield  {author} {\bibinfo {author} {\bibfnamefont {S.}~\bibnamefont
			{Kim}}, \bibinfo {author} {\bibfnamefont {C.}~\bibnamefont {Laschi}},\ and\
		\bibinfo {author} {\bibfnamefont {B.}~\bibnamefont {Trimmer}},\ }\bibfield
	{title} {\bibinfo {title} {{Soft robotics: A bioinspired evolution in
				robotics}},\ }\href@noop {} {\bibfield  {journal} {\bibinfo  {journal}
			{Trends in Biotechnology}\ }\textbf {\bibinfo {volume} {31}},\ \bibinfo
		{pages} {287} (\bibinfo {year} {2013})}\BibitemShut {NoStop}%
	\bibitem [{\citenamefont {Friedrich}\ and\ \citenamefont
		{J{\"{u}}licher}(2007)}]{Friedrich2007}%
	\BibitemOpen
	\bibfield  {author} {\bibinfo {author} {\bibfnamefont {B.~M.}\ \bibnamefont
			{Friedrich}}\ and\ \bibinfo {author} {\bibfnamefont {F.}~\bibnamefont
			{J{\"{u}}licher}},\ }\bibfield  {title} {\bibinfo {title} {{Chemotaxis of
				sperm cells}},\ }\href {https://doi.org/10.1073/pnas.0703530104} {\bibfield
		{journal} {\bibinfo  {journal} {Proceedings of the National Academy of
				Sciences of the United States of America}\ }\textbf {\bibinfo {volume}
			{104}},\ \bibinfo {pages} {13256} (\bibinfo {year} {2007})}\BibitemShut
	{NoStop}%
	\bibitem [{\citenamefont {Kirkman-Brown}\ and\ \citenamefont
		{Smith}(2011)}]{Kirkman-Brown2011}%
	\BibitemOpen
	\bibfield  {author} {\bibinfo {author} {\bibfnamefont {J.~C.}\ \bibnamefont
			{Kirkman-Brown}}\ and\ \bibinfo {author} {\bibfnamefont {D.~J.}\ \bibnamefont
			{Smith}},\ }\bibfield  {title} {\bibinfo {title} {{Sperm motility: Is
				viscosity fundamental to progress?}},\ }\bibfield  {booktitle} {\emph
		{\bibinfo {booktitle} {Molecular Human Reproduction}},\ }\href@noop {}
	{\bibfield  {journal} {\bibinfo  {journal} {Molecular Human Reproduction}\
		}\textbf {\bibinfo {volume} {17}},\ \bibinfo {pages} {539} (\bibinfo {year}
		{2011})}\BibitemShut {NoStop}%
	\bibitem [{\citenamefont {Tung}\ \emph {et~al.}(2017)\citenamefont {Tung},
		\citenamefont {Lin}, \citenamefont {Harvey}, \citenamefont {Fiore},
		\citenamefont {Ardon}, \citenamefont {Wu},\ and\ \citenamefont
		{Suarez}}]{Tung2017}%
	\BibitemOpen
	\bibfield  {author} {\bibinfo {author} {\bibfnamefont {C.~K.}\ \bibnamefont
			{Tung}}, \bibinfo {author} {\bibfnamefont {C.}~\bibnamefont {Lin}}, \bibinfo
		{author} {\bibfnamefont {B.}~\bibnamefont {Harvey}}, \bibinfo {author}
		{\bibfnamefont {A.~G.}\ \bibnamefont {Fiore}}, \bibinfo {author}
		{\bibfnamefont {F.}~\bibnamefont {Ardon}}, \bibinfo {author} {\bibfnamefont
			{M.}~\bibnamefont {Wu}},\ and\ \bibinfo {author} {\bibfnamefont {S.~S.}\
			\bibnamefont {Suarez}},\ }\bibfield  {title} {\bibinfo {title} {{Fluid
				viscoelasticity promotes collective swimming of sperm}},\ }\href
	{https://doi.org/10.1038/s41598-017-03341-4} {\bibfield  {journal} {\bibinfo
			{journal} {Scientific Reports}\ }\textbf {\bibinfo {volume} {7}},\ \bibinfo
		{pages} {1} (\bibinfo {year} {2017})}\BibitemShut {NoStop}%
	\bibitem [{\citenamefont {Harman}\ \emph {et~al.}(2012)\citenamefont {Harman},
		\citenamefont {Dunham-Ems}, \citenamefont {Caimano}, \citenamefont
		{Belperron}, \citenamefont {Bockenstedt}, \citenamefont {Fu}, \citenamefont
		{Radolf},\ and\ \citenamefont {Wolgemuth}}]{Harman2012}%
	\BibitemOpen
	\bibfield  {author} {\bibinfo {author} {\bibfnamefont {M.~W.}\ \bibnamefont
			{Harman}}, \bibinfo {author} {\bibfnamefont {S.~M.}\ \bibnamefont
			{Dunham-Ems}}, \bibinfo {author} {\bibfnamefont {M.~J.}\ \bibnamefont
			{Caimano}}, \bibinfo {author} {\bibfnamefont {A.~A.}\ \bibnamefont
			{Belperron}}, \bibinfo {author} {\bibfnamefont {L.~K.}\ \bibnamefont
			{Bockenstedt}}, \bibinfo {author} {\bibfnamefont {H.~C.}\ \bibnamefont {Fu}},
		\bibinfo {author} {\bibfnamefont {J.~D.}\ \bibnamefont {Radolf}},\ and\
		\bibinfo {author} {\bibfnamefont {C.~W.}\ \bibnamefont {Wolgemuth}},\
	}\bibfield  {title} {\bibinfo {title} {{The heterogeneous motility of the
				Lyme disease spirochete in gelatin mimics dissemination through tissue}},\
	}\href {https://doi.org/10.1073/pnas.1114362109} {\bibfield  {journal}
		{\bibinfo  {journal} {Proceedings of the National Academy of Sciences of the
				United States of America}\ }\textbf {\bibinfo {volume} {109}},\ \bibinfo
		{pages} {3059} (\bibinfo {year} {2012})}\BibitemShut {NoStop}%
	\bibitem [{\citenamefont {Figueroa-Morales}\ \emph {et~al.}(2019)\citenamefont
		{Figueroa-Morales}, \citenamefont {Dominguez-Rubio}, \citenamefont {Ott},\
		and\ \citenamefont {Aranson}}]{Figueroa-Morales2019}%
	\BibitemOpen
	\bibfield  {author} {\bibinfo {author} {\bibfnamefont {N.}~\bibnamefont
			{Figueroa-Morales}}, \bibinfo {author} {\bibfnamefont {L.}~\bibnamefont
			{Dominguez-Rubio}}, \bibinfo {author} {\bibfnamefont {T.~L.}\ \bibnamefont
			{Ott}},\ and\ \bibinfo {author} {\bibfnamefont {I.~S.}\ \bibnamefont
			{Aranson}},\ }\bibfield  {title} {\bibinfo {title} {{Mechanical shear
				controls bacterial penetration in mucus}},\ }\href
	{/pmc/articles/PMC6609767/?report=abstract
		https://www.ncbi.nlm.nih.gov/pmc/articles/PMC6609767/} {\bibfield  {journal}
		{\bibinfo  {journal} {Scientific Reports}\ }\textbf {\bibinfo {volume} {9}}
		(\bibinfo {year} {2019})}\BibitemShut {NoStop}%
	\bibitem [{\citenamefont {Fu}\ \emph {et~al.}(2007)\citenamefont {Fu},
		\citenamefont {Powers},\ and\ \citenamefont {Wolgemuth}}]{Fu2007a}%
	\BibitemOpen
	\bibfield  {author} {\bibinfo {author} {\bibfnamefont {H.~C.}\ \bibnamefont
			{Fu}}, \bibinfo {author} {\bibfnamefont {T.~R.}\ \bibnamefont {Powers}},\
		and\ \bibinfo {author} {\bibfnamefont {C.~W.}\ \bibnamefont {Wolgemuth}},\
	}\bibfield  {title} {\bibinfo {title} {{Theory of swimming filaments in
				viscoelastic media}},\ }\href@noop {} {\bibfield  {journal} {\bibinfo
			{journal} {Physical Review Letters}\ }\textbf {\bibinfo {volume} {99}},\
		\bibinfo {pages} {258101} (\bibinfo {year} {2007})}\BibitemShut {NoStop}%
	\bibitem [{\citenamefont {Jabbarzadeh}\ \emph {et~al.}(2014)\citenamefont
		{Jabbarzadeh}, \citenamefont {Hyon},\ and\ \citenamefont
		{Fu}}]{Jabbarzadeh2014}%
	\BibitemOpen
	\bibfield  {author} {\bibinfo {author} {\bibfnamefont {M.}~\bibnamefont
			{Jabbarzadeh}}, \bibinfo {author} {\bibfnamefont {Y.}~\bibnamefont {Hyon}},\
		and\ \bibinfo {author} {\bibfnamefont {H.~C.}\ \bibnamefont {Fu}},\
	}\bibfield  {title} {\bibinfo {title} {{Swimming fluctuations of
				micro-organisms due to heterogeneous microstructure}},\ }\href
	{https://doi.org/10.1103/PhysRevE.90.043021} {\bibfield  {journal} {\bibinfo
			{journal} {Physical Review E}\ }\textbf {\bibinfo {volume} {90}},\ \bibinfo
		{pages} {043021} (\bibinfo {year} {2014})}\BibitemShut {NoStop}%
	\bibitem [{\citenamefont {Martinez}\ \emph {et~al.}(2014)\citenamefont
		{Martinez}, \citenamefont {Schwarz-Linek}, \citenamefont {Reufer},
		\citenamefont {Wilson}, \citenamefont {Morozov},\ and\ \citenamefont
		{Poon}}]{Martinez2014}%
	\BibitemOpen
	\bibfield  {author} {\bibinfo {author} {\bibfnamefont {V.~A.}\ \bibnamefont
			{Martinez}}, \bibinfo {author} {\bibfnamefont {J.}~\bibnamefont
			{Schwarz-Linek}}, \bibinfo {author} {\bibfnamefont {M.}~\bibnamefont
			{Reufer}}, \bibinfo {author} {\bibfnamefont {L.~G.}\ \bibnamefont {Wilson}},
		\bibinfo {author} {\bibfnamefont {A.~N.}\ \bibnamefont {Morozov}},\ and\
		\bibinfo {author} {\bibfnamefont {W.~C.}\ \bibnamefont {Poon}},\ }\bibfield
	{title} {\bibinfo {title} {{Flagellated bacterial motility in polymer
				solutions}},\ }\href@noop {} {\bibfield  {journal} {\bibinfo  {journal}
			{Proceedings of the National Academy of Sciences of the United States of
				America}\ }\textbf {\bibinfo {volume} {111}},\ \bibinfo {pages} {17771}
		(\bibinfo {year} {2014})}\BibitemShut {NoStop}%
	\bibitem [{\citenamefont {Wang}\ \emph {et~al.}(2019)\citenamefont {Wang},
		\citenamefont {Wu}, \citenamefont {Lin}, \citenamefont {Si},\ and\
		\citenamefont {He}}]{Wang2019}%
	\BibitemOpen
	\bibfield  {author} {\bibinfo {author} {\bibfnamefont {W.}~\bibnamefont
			{Wang}}, \bibinfo {author} {\bibfnamefont {Z.}~\bibnamefont {Wu}}, \bibinfo
		{author} {\bibfnamefont {X.}~\bibnamefont {Lin}}, \bibinfo {author}
		{\bibfnamefont {T.}~\bibnamefont {Si}},\ and\ \bibinfo {author}
		{\bibfnamefont {Q.}~\bibnamefont {He}},\ }\bibfield  {title} {\bibinfo
		{title} {{Gold-Nanoshell-Functionalized Polymer Nanoswimmer for
				Photomechanical Poration of Single-Cell Membrane}},\ }\href
	{https://doi.org/10.1021/jacs.8b13882} {\bibfield  {journal} {\bibinfo
			{journal} {Journal of the American Chemical Society}\ }\textbf {\bibinfo
			{volume} {141}},\ \bibinfo {pages} {6601} (\bibinfo {year}
		{2019})}\BibitemShut {NoStop}%
	\bibitem [{\citenamefont {Guillamat}\ \emph {et~al.}(2017)\citenamefont
		{Guillamat}, \citenamefont {Ign{\'{e}}s-Mullol},\ and\ \citenamefont
		{Sagu{\'{e}}s}}]{Guillamat2017}%
	\BibitemOpen
	\bibfield  {author} {\bibinfo {author} {\bibfnamefont {P.}~\bibnamefont
			{Guillamat}}, \bibinfo {author} {\bibfnamefont {J.}~\bibnamefont
			{Ign{\'{e}}s-Mullol}},\ and\ \bibinfo {author} {\bibfnamefont
			{F.}~\bibnamefont {Sagu{\'{e}}s}},\ }\bibfield  {title} {\bibinfo {title}
		{{Taming active turbulence with patterned soft interfaces}},\ }\href
	{www.nature.com/naturecommunications} {\bibfield  {journal} {\bibinfo
			{journal} {Nature Communications}\ }\textbf {\bibinfo {volume} {8}},\
		\bibinfo {pages} {1} (\bibinfo {year} {2017})}\BibitemShut {NoStop}%
	\bibitem [{\citenamefont {Aranson}(2018)}]{Aranson2018}%
	\BibitemOpen
	\bibfield  {author} {\bibinfo {author} {\bibfnamefont {I.~S.}\ \bibnamefont
			{Aranson}},\ }\bibfield  {title} {\bibinfo {title} {{Harnessing Medium
				Anisotropy to Control Active Matter}},\ }\href
	{https://doi.org/10.1021/acs.accounts.8b00300} {\bibfield  {journal}
		{\bibinfo  {journal} {Accounts of Chemical Research}\ }\textbf {\bibinfo
			{volume} {51}},\ \bibinfo {pages} {3023} (\bibinfo {year}
		{2018})}\BibitemShut {NoStop}%
	\bibitem [{\citenamefont {Juarez}\ \emph {et~al.}(2010)\citenamefont {Juarez},
		\citenamefont {Lu}, \citenamefont {Sznitman},\ and\ \citenamefont
		{Arratia}}]{Juarez2010}%
	\BibitemOpen
	\bibfield  {author} {\bibinfo {author} {\bibfnamefont {G.}~\bibnamefont
			{Juarez}}, \bibinfo {author} {\bibfnamefont {K.}~\bibnamefont {Lu}}, \bibinfo
		{author} {\bibfnamefont {J.}~\bibnamefont {Sznitman}},\ and\ \bibinfo
		{author} {\bibfnamefont {P.~E.}\ \bibnamefont {Arratia}},\ }\bibfield
	{title} {\bibinfo {title} {{Motility of small nematodes in wet granular
				media}},\ }\href {https://doi.org/10.1209/0295-5075/92/44002} {\bibfield
		{journal} {\bibinfo  {journal} {Europhysics Letters}\ }\textbf {\bibinfo
			{volume} {92}},\ \bibinfo {pages} {44002} (\bibinfo {year}
		{2010})}\BibitemShut {NoStop}%
	\bibitem [{\citenamefont {Lauga}\ \emph {et~al.}(2006)\citenamefont {Lauga},
		\citenamefont {DiLuzio}, \citenamefont {Whitesides},\ and\ \citenamefont
		{Stone}}]{Lauga2006}%
	\BibitemOpen
	\bibfield  {author} {\bibinfo {author} {\bibfnamefont {E.}~\bibnamefont
			{Lauga}}, \bibinfo {author} {\bibfnamefont {W.~R.}\ \bibnamefont {DiLuzio}},
		\bibinfo {author} {\bibfnamefont {G.~M.}\ \bibnamefont {Whitesides}},\ and\
		\bibinfo {author} {\bibfnamefont {H.~A.}\ \bibnamefont {Stone}},\ }\bibfield
	{title} {\bibinfo {title} {{Swimming in circles: Motion of bacteria near
				solid boundaries}},\ }\href {https://doi.org/10.1529/biophysj.105.069401}
	{\bibfield  {journal} {\bibinfo  {journal} {Biophysical Journal}\ }\textbf
		{\bibinfo {volume} {90}},\ \bibinfo {pages} {400} (\bibinfo {year}
		{2006})}\BibitemShut {NoStop}%
	\bibitem [{\citenamefont {Molaei}\ and\ \citenamefont
		{Sheng}(2016)}]{Molaei2016}%
	\BibitemOpen
	\bibfield  {author} {\bibinfo {author} {\bibfnamefont {M.}~\bibnamefont
			{Molaei}}\ and\ \bibinfo {author} {\bibfnamefont {J.}~\bibnamefont {Sheng}},\
	}\bibfield  {title} {\bibinfo {title} {{Succeed escape: Flow shear promotes
				tumbling of Escherichia colinear a solid surface}},\ }\href
	{https://doi.org/10.1038/srep35290} {\bibfield  {journal} {\bibinfo
			{journal} {Scientific Reports}\ }\textbf {\bibinfo {volume} {6}},\ \bibinfo
		{pages} {1} (\bibinfo {year} {2016})}\BibitemShut {NoStop}%
	\bibitem [{\citenamefont {Berke}\ \emph {et~al.}(2008)\citenamefont {Berke},
		\citenamefont {Turner}, \citenamefont {Berg},\ and\ \citenamefont
		{Lauga}}]{Berke2008}%
	\BibitemOpen
	\bibfield  {author} {\bibinfo {author} {\bibfnamefont {A.~P.}\ \bibnamefont
			{Berke}}, \bibinfo {author} {\bibfnamefont {L.}~\bibnamefont {Turner}},
		\bibinfo {author} {\bibfnamefont {H.~C.}\ \bibnamefont {Berg}},\ and\
		\bibinfo {author} {\bibfnamefont {E.}~\bibnamefont {Lauga}},\ }\bibfield
	{title} {\bibinfo {title} {{Hydrodynamic attraction of swimming
				microorganisms by surfaces}},\ }\href
	{https://journals.aps.org/prl/abstract/10.1103/PhysRevLett.101.038102}
	{\bibfield  {journal} {\bibinfo  {journal} {Physical Review Letters}\
		}\textbf {\bibinfo {volume} {101}},\ \bibinfo {pages} {038102} (\bibinfo
		{year} {2008})}\BibitemShut {NoStop}%
	\bibitem [{\citenamefont {Li}\ and\ \citenamefont {Tang}(2009)}]{Li2009}%
	\BibitemOpen
	\bibfield  {author} {\bibinfo {author} {\bibfnamefont {G.}~\bibnamefont
			{Li}}\ and\ \bibinfo {author} {\bibfnamefont {J.~X.}\ \bibnamefont {Tang}},\
	}\bibfield  {title} {\bibinfo {title} {{Accumulation of microswimmers near a
				surface mediated by collision and rotational Brownian motion}},\ }\href
	{https://doi.org/10.1103/PhysRevLett.103.078101} {\bibfield  {journal}
		{\bibinfo  {journal} {Physical Review Letters}\ }\textbf {\bibinfo {volume}
			{103}},\ \bibinfo {pages} {078101} (\bibinfo {year} {2009})}\BibitemShut
	{NoStop}%
	\bibitem [{\citenamefont {Deng}\ \emph {et~al.}(2020)\citenamefont {Deng},
		\citenamefont {Molaei}, \citenamefont {Chisholm},\ and\ \citenamefont
		{Stebe}}]{Deng2020}%
	\BibitemOpen
	\bibfield  {author} {\bibinfo {author} {\bibfnamefont {J.}~\bibnamefont
			{Deng}}, \bibinfo {author} {\bibfnamefont {M.}~\bibnamefont {Molaei}},
		\bibinfo {author} {\bibfnamefont {N.~G.}\ \bibnamefont {Chisholm}},\ and\
		\bibinfo {author} {\bibfnamefont {K.~J.}\ \bibnamefont {Stebe}},\ }\bibfield
	{title} {\bibinfo {title} {{Motile Bacteria at Oil-Water Interfaces:
				Pseudomonas aeruginosa}},\ }\href
	{https://doi.org/10.1021/acs.langmuir.9b03578} {\bibfield  {journal}
		{\bibinfo  {journal} {Langmuir}\ }\textbf {\bibinfo {volume} {36}},\ \bibinfo
		{pages} {6888} (\bibinfo {year} {2020})}\BibitemShut {NoStop}%
	\bibitem [{\citenamefont {Hu}\ \emph {et~al.}(2015)\citenamefont {Hu},
		\citenamefont {Wysocki}, \citenamefont {Winkler},\ and\ \citenamefont
		{Gompper}}]{Hu2015}%
	\BibitemOpen
	\bibfield  {author} {\bibinfo {author} {\bibfnamefont {J.}~\bibnamefont
			{Hu}}, \bibinfo {author} {\bibfnamefont {A.}~\bibnamefont {Wysocki}},
		\bibinfo {author} {\bibfnamefont {R.~G.}\ \bibnamefont {Winkler}},\ and\
		\bibinfo {author} {\bibfnamefont {G.}~\bibnamefont {Gompper}},\ }\bibfield
	{title} {\bibinfo {title} {{Physical sensing of surface properties by
				microswimmers-directing bacterial motion via wall slip}},\ }\href
	{https://doi.org/10.1038/srep09586} {\bibfield  {journal} {\bibinfo
			{journal} {Scientific Reports}\ }\textbf {\bibinfo {volume} {5}},\ \bibinfo
		{pages} {1} (\bibinfo {year} {2015})}\BibitemShut {NoStop}%
	\bibitem [{\citenamefont {Desai}\ \emph {et~al.}(2018)\citenamefont {Desai},
		\citenamefont {Shaik},\ and\ \citenamefont {Ardekani}}]{Desai2018}%
	\BibitemOpen
	\bibfield  {author} {\bibinfo {author} {\bibfnamefont {N.}~\bibnamefont
			{Desai}}, \bibinfo {author} {\bibfnamefont {V.~A.}\ \bibnamefont {Shaik}},\
		and\ \bibinfo {author} {\bibfnamefont {A.~M.}\ \bibnamefont {Ardekani}},\
	}\bibfield  {title} {\bibinfo {title} {{Hydrodynamics-mediated trapping of
				micro-swimmers near drops}},\ }\href {https://doi.org/10.1039/c7sm01615h}
	{\bibfield  {journal} {\bibinfo  {journal} {Soft Matter}\ }\textbf {\bibinfo
			{volume} {14}},\ \bibinfo {pages} {264} (\bibinfo {year} {2018})}\BibitemShut
	{NoStop}%
	\bibitem [{\citenamefont {Prakash}\ \emph {et~al.}(2020)\citenamefont
		{Prakash}, \citenamefont {Abdulla}, \citenamefont {Singh},\ and\
		\citenamefont {Varma}}]{Prakash2020}%
	\BibitemOpen
	\bibfield  {author} {\bibinfo {author} {\bibfnamefont {P.}~\bibnamefont
			{Prakash}}, \bibinfo {author} {\bibfnamefont {A.~Z.}\ \bibnamefont
			{Abdulla}}, \bibinfo {author} {\bibfnamefont {V.}~\bibnamefont {Singh}},\
		and\ \bibinfo {author} {\bibfnamefont {M.}~\bibnamefont {Varma}},\ }\bibfield
	{title} {\bibinfo {title} {{Swimming statistics of cargo-loaded single
				bacteria}},\ }\href {https://doi.org/10.1039/d0sm01066a} {\bibfield
		{journal} {\bibinfo  {journal} {Soft Matter}\ }\textbf {\bibinfo {volume}
			{16}},\ \bibinfo {pages} {9499} (\bibinfo {year} {2020})}\BibitemShut
	{NoStop}%
	\bibitem [{\citenamefont {Fernandez}\ \emph {et~al.}(2019)\citenamefont
		{Fernandez}, \citenamefont {Stocker},\ and\ \citenamefont
		{Juarez}}]{Fernandez2019}%
	\BibitemOpen
	\bibfield  {author} {\bibinfo {author} {\bibfnamefont {V.~I.}\ \bibnamefont
			{Fernandez}}, \bibinfo {author} {\bibfnamefont {R.}~\bibnamefont {Stocker}},\
		and\ \bibinfo {author} {\bibfnamefont {G.}~\bibnamefont {Juarez}},\
	}\bibfield  {title} {\bibinfo {title} {{Modeling the Impact of Dilution on
				the Microbial Degradation of Dispersed Oil in Marine Environments}},\ }in\
	\href {https://doi.org/10.1201/9781315164700-13} {\emph {\bibinfo {booktitle}
			{Oilfield Microbiology}}},\ \bibinfo {editor} {edited by\ \bibinfo {editor}
		{\bibfnamefont {T.~L.}\ \bibnamefont {Skovhus}}\ and\ \bibinfo {editor}
		{\bibfnamefont {C.}~\bibnamefont {Whitby}}}\ (\bibinfo  {publisher} {CRC
		Press},\ \bibinfo {year} {2019})\ pp.\ \bibinfo {pages}
	{215--232}\BibitemShut {NoStop}%
	\bibitem [{\citenamefont {Galajda}\ \emph {et~al.}(2007)\citenamefont
		{Galajda}, \citenamefont {Keymer}, \citenamefont {Chaikin},\ and\
		\citenamefont {Austin}}]{Galajda2007}%
	\BibitemOpen
	\bibfield  {author} {\bibinfo {author} {\bibfnamefont {P.}~\bibnamefont
			{Galajda}}, \bibinfo {author} {\bibfnamefont {J.}~\bibnamefont {Keymer}},
		\bibinfo {author} {\bibfnamefont {P.}~\bibnamefont {Chaikin}},\ and\ \bibinfo
		{author} {\bibfnamefont {R.}~\bibnamefont {Austin}},\ }\bibfield  {title}
	{\bibinfo {title} {{A wall of funnels concentrates swimming bacteria}},\
	}\href@noop {} {\bibfield  {journal} {\bibinfo  {journal} {Journal of
				Bacteriology}\ }\textbf {\bibinfo {volume} {189}},\ \bibinfo {pages} {8704}
		(\bibinfo {year} {2007})}\BibitemShut {NoStop}%
	\bibitem [{\citenamefont {Austin}\ \emph {et~al.}(2017)\citenamefont {Austin},
		\citenamefont {Caro}, \citenamefont {Sankar}, \citenamefont {Penniman},
		\citenamefont {Perdomo}, \citenamefont {Hu}, \citenamefont {Patel},
		\citenamefont {Gu}, \citenamefont {Watve}, \citenamefont {Hammer},\ and\
		\citenamefont {Forest}}]{Austin2017}%
	\BibitemOpen
	\bibfield  {author} {\bibinfo {author} {\bibfnamefont {C.~M.}\ \bibnamefont
			{Austin}}, \bibinfo {author} {\bibfnamefont {D.~M.}\ \bibnamefont {Caro}},
		\bibinfo {author} {\bibfnamefont {S.}~\bibnamefont {Sankar}}, \bibinfo
		{author} {\bibfnamefont {W.~F.}\ \bibnamefont {Penniman}}, \bibinfo {author}
		{\bibfnamefont {J.~E.}\ \bibnamefont {Perdomo}}, \bibinfo {author}
		{\bibfnamefont {L.}~\bibnamefont {Hu}}, \bibinfo {author} {\bibfnamefont
			{S.}~\bibnamefont {Patel}}, \bibinfo {author} {\bibfnamefont
			{X.}~\bibnamefont {Gu}}, \bibinfo {author} {\bibfnamefont {S.}~\bibnamefont
			{Watve}}, \bibinfo {author} {\bibfnamefont {B.~K.}\ \bibnamefont {Hammer}},\
		and\ \bibinfo {author} {\bibfnamefont {C.~R.}\ \bibnamefont {Forest}},\
	}\bibfield  {title} {\bibinfo {title} {{Porous monolith microfluidics for
				bacterial cell-to-cell communication assays}},\ }\href
	{https://doi.org/10.1063/1.4995597} {\bibfield  {journal} {\bibinfo
			{journal} {Biomicrofluidics}\ }\textbf {\bibinfo {volume} {11}},\ \bibinfo
		{pages} {044110} (\bibinfo {year} {2017})}\BibitemShut {NoStop}%
	\bibitem [{\citenamefont {Bianchi}\ \emph {et~al.}(2017)\citenamefont
		{Bianchi}, \citenamefont {Saglimbeni},\ and\ \citenamefont {{Di
				Leonardo}}}]{Bianchi2017}%
	\BibitemOpen
	\bibfield  {author} {\bibinfo {author} {\bibfnamefont {S.}~\bibnamefont
			{Bianchi}}, \bibinfo {author} {\bibfnamefont {F.}~\bibnamefont
			{Saglimbeni}},\ and\ \bibinfo {author} {\bibfnamefont {R.}~\bibnamefont {{Di
					Leonardo}}},\ }\bibfield  {title} {\bibinfo {title} {{Holographic imaging
				reveals the mechanism of wall entrapment in swimming bacteria}},\ }\href
	{https://doi.org/10.1103/PhysRevX.7.011010} {\bibfield  {journal} {\bibinfo
			{journal} {Physical Review X}\ }\textbf {\bibinfo {volume} {7}},\ \bibinfo
		{pages} {011010} (\bibinfo {year} {2017})}\BibitemShut {NoStop}%
	\bibitem [{\citenamefont {Secchi}\ \emph {et~al.}(2020)\citenamefont {Secchi},
		\citenamefont {Vitale}, \citenamefont {Mi{\~{n}}o}, \citenamefont {Kantsler},
		\citenamefont {Eberl}, \citenamefont {Rusconi},\ and\ \citenamefont
		{Stocker}}]{Secchi2020}%
	\BibitemOpen
	\bibfield  {author} {\bibinfo {author} {\bibfnamefont {E.}~\bibnamefont
			{Secchi}}, \bibinfo {author} {\bibfnamefont {A.}~\bibnamefont {Vitale}},
		\bibinfo {author} {\bibfnamefont {G.~L.}\ \bibnamefont {Mi{\~{n}}o}},
		\bibinfo {author} {\bibfnamefont {V.}~\bibnamefont {Kantsler}}, \bibinfo
		{author} {\bibfnamefont {L.}~\bibnamefont {Eberl}}, \bibinfo {author}
		{\bibfnamefont {R.}~\bibnamefont {Rusconi}},\ and\ \bibinfo {author}
		{\bibfnamefont {R.}~\bibnamefont {Stocker}},\ }\bibfield  {title} {\bibinfo
		{title} {{The effect of flow on swimming bacteria controls the initial
				colonization of curved surfaces}},\ }\href
	{https://doi.org/10.1038/s41467-020-16620-y} {\bibfield  {journal} {\bibinfo
			{journal} {Nature Communications}\ }\textbf {\bibinfo {volume} {11}},\
		\bibinfo {pages} {1} (\bibinfo {year} {2020})}\BibitemShut {NoStop}%
	\bibitem [{\citenamefont {Denissenko}\ \emph {et~al.}(2012)\citenamefont
		{Denissenko}, \citenamefont {Kantsler}, \citenamefont {Smith},\ and\
		\citenamefont {Kirkman-Brown}}]{Denissenko2012}%
	\BibitemOpen
	\bibfield  {author} {\bibinfo {author} {\bibfnamefont {P.}~\bibnamefont
			{Denissenko}}, \bibinfo {author} {\bibfnamefont {V.}~\bibnamefont
			{Kantsler}}, \bibinfo {author} {\bibfnamefont {D.~J.}\ \bibnamefont
			{Smith}},\ and\ \bibinfo {author} {\bibfnamefont {J.}~\bibnamefont
			{Kirkman-Brown}},\ }\bibfield  {title} {\bibinfo {title} {{Human spermatozoa
				migration in microchannels reveals boundary-following navigation}},\ }\href
	{https://doi.org/10.1073/pnas.1202934109} {\bibfield  {journal} {\bibinfo
			{journal} {Proceedings of the National Academy of Sciences of the United
				States of America}\ }\textbf {\bibinfo {volume} {109}},\ \bibinfo {pages}
		{8007} (\bibinfo {year} {2012})}\BibitemShut {NoStop}%
	\bibitem [{\citenamefont {Kantsler}\ \emph {et~al.}(2014)\citenamefont
		{Kantsler}, \citenamefont {Dunkel}, \citenamefont {Blayney},\ and\
		\citenamefont {Goldstein}}]{Kantsler2014}%
	\BibitemOpen
	\bibfield  {author} {\bibinfo {author} {\bibfnamefont {V.}~\bibnamefont
			{Kantsler}}, \bibinfo {author} {\bibfnamefont {J.}~\bibnamefont {Dunkel}},
		\bibinfo {author} {\bibfnamefont {M.}~\bibnamefont {Blayney}},\ and\ \bibinfo
		{author} {\bibfnamefont {R.~E.}\ \bibnamefont {Goldstein}},\ }\bibfield
	{title} {\bibinfo {title} {{Rheotaxis facilitates upstream navigation of
				mammalian sperm cells}},\ }\href@noop {} {\bibfield  {journal} {\bibinfo
			{journal} {eLife}\ }\textbf {\bibinfo {volume} {2014}} (\bibinfo {year}
		{2014})}\BibitemShut {NoStop}%
	\bibitem [{\citenamefont {Rode}\ \emph {et~al.}(2019)\citenamefont {Rode},
		\citenamefont {Elgeti},\ and\ \citenamefont {Gompper}}]{Rode2019}%
	\BibitemOpen
	\bibfield  {author} {\bibinfo {author} {\bibfnamefont {S.}~\bibnamefont
			{Rode}}, \bibinfo {author} {\bibfnamefont {J.}~\bibnamefont {Elgeti}},\ and\
		\bibinfo {author} {\bibfnamefont {G.}~\bibnamefont {Gompper}},\ }\bibfield
	{title} {\bibinfo {title} {{Sperm motility in modulated microchannels}},\
	}\href {https://doi.org/10.1088/1367-2630/aaf544} {\bibfield  {journal}
		{\bibinfo  {journal} {New Journal of Physics}\ }\textbf {\bibinfo {volume}
			{21}},\ \bibinfo {pages} {013016} (\bibinfo {year} {2019})}\BibitemShut
	{NoStop}%
	\bibitem [{\citenamefont {K{\"{u}}hn}\ \emph {et~al.}(2017)\citenamefont
		{K{\"{u}}hn}, \citenamefont {Schmidt}, \citenamefont {Eckhardt},\ and\
		\citenamefont {Thormann}}]{Kuhn2017}%
	\BibitemOpen
	\bibfield  {author} {\bibinfo {author} {\bibfnamefont {M.~J.}\ \bibnamefont
			{K{\"{u}}hn}}, \bibinfo {author} {\bibfnamefont {F.~K.}\ \bibnamefont
			{Schmidt}}, \bibinfo {author} {\bibfnamefont {B.}~\bibnamefont {Eckhardt}},\
		and\ \bibinfo {author} {\bibfnamefont {K.~M.}\ \bibnamefont {Thormann}},\
	}\bibfield  {title} {\bibinfo {title} {{Bacteria exploit a polymorphic
				instability of the flagellar filament to escape from traps}},\ }\href
	{https://doi.org/10.1073/pnas.1701644114} {\bibfield  {journal} {\bibinfo
			{journal} {Proceedings of the National Academy of Sciences of the United
				States of America}\ }\textbf {\bibinfo {volume} {114}},\ \bibinfo {pages}
		{6340} (\bibinfo {year} {2017})}\BibitemShut {NoStop}%
	\bibitem [{\citenamefont {Bhattacharjee}\ and\ \citenamefont
		{Datta}(2019)}]{Bhattacharjee2019}%
	\BibitemOpen
	\bibfield  {author} {\bibinfo {author} {\bibfnamefont {T.}~\bibnamefont
			{Bhattacharjee}}\ and\ \bibinfo {author} {\bibfnamefont {S.~S.}\ \bibnamefont
			{Datta}},\ }\bibfield  {title} {\bibinfo {title} {{Bacterial hopping and
				trapping in porous media}},\ }\href
	{https://doi.org/10.1038/s41467-019-10115-1} {\bibfield  {journal} {\bibinfo
			{journal} {Nature Communications}\ }\textbf {\bibinfo {volume} {10}},\
		\bibinfo {pages} {1} (\bibinfo {year} {2019})}\BibitemShut {NoStop}%
	\bibitem [{\citenamefont {Jenkinson}\ \emph {et~al.}(2018)\citenamefont
		{Jenkinson}, \citenamefont {Seuront}, \citenamefont {DIng},\ and\
		\citenamefont {Elias}}]{Jenkinson2018}%
	\BibitemOpen
	\bibfield  {author} {\bibinfo {author} {\bibfnamefont {I.~R.}\ \bibnamefont
			{Jenkinson}}, \bibinfo {author} {\bibfnamefont {L.}~\bibnamefont {Seuront}},
		\bibinfo {author} {\bibfnamefont {H.}~\bibnamefont {DIng}},\ and\ \bibinfo
		{author} {\bibfnamefont {F.}~\bibnamefont {Elias}},\ }\bibfield  {title}
	{\bibinfo {title} {{Biological modification of mechanical properties of the
				sea surface microlayer, influencing waves, ripples, foam and air-sea
				fluxes}},\ }\href@noop {} {\bibfield  {journal} {\bibinfo  {journal}
			{Elementa: Science of the Anthropocene}\ }\textbf {\bibinfo {volume} {6}}
		(\bibinfo {year} {2018})}\BibitemShut {NoStop}%
	\bibitem [{\citenamefont {Majmudar}\ \emph {et~al.}(2012)\citenamefont
		{Majmudar}, \citenamefont {Keaveny}, \citenamefont {Zhang},\ and\
		\citenamefont {Shelley}}]{Majmudar2012}%
	\BibitemOpen
	\bibfield  {author} {\bibinfo {author} {\bibfnamefont {T.}~\bibnamefont
			{Majmudar}}, \bibinfo {author} {\bibfnamefont {E.~E.}\ \bibnamefont
			{Keaveny}}, \bibinfo {author} {\bibfnamefont {J.}~\bibnamefont {Zhang}},\
		and\ \bibinfo {author} {\bibfnamefont {M.~J.}\ \bibnamefont {Shelley}},\
	}\bibfield  {title} {\bibinfo {title} {{Experiments and theory of undulatory
				locomotion in a simple structured medium}},\ }\href@noop {} {\bibfield
		{journal} {\bibinfo  {journal} {Journal of The Royal Society Interface}\
		}\textbf {\bibinfo {volume} {9}},\ \bibinfo {pages} {1809} (\bibinfo {year}
		{2012})}\BibitemShut {NoStop}%
	\bibitem [{\citenamefont {Li}\ \emph {et~al.}(2014)\citenamefont {Li},
		\citenamefont {Qiu}, \citenamefont {Glidle}, \citenamefont {McIlvenna},
		\citenamefont {Luo}, \citenamefont {Cooper}, \citenamefont {Shi},\ and\
		\citenamefont {Yin}}]{Li2014}%
	\BibitemOpen
	\bibfield  {author} {\bibinfo {author} {\bibfnamefont {B.}~\bibnamefont
			{Li}}, \bibinfo {author} {\bibfnamefont {Y.}~\bibnamefont {Qiu}}, \bibinfo
		{author} {\bibfnamefont {A.}~\bibnamefont {Glidle}}, \bibinfo {author}
		{\bibfnamefont {D.}~\bibnamefont {McIlvenna}}, \bibinfo {author}
		{\bibfnamefont {Q.}~\bibnamefont {Luo}}, \bibinfo {author} {\bibfnamefont
			{J.}~\bibnamefont {Cooper}}, \bibinfo {author} {\bibfnamefont {H.~C.}\
			\bibnamefont {Shi}},\ and\ \bibinfo {author} {\bibfnamefont {H.}~\bibnamefont
			{Yin}},\ }\bibfield  {title} {\bibinfo {title} {{Gradient microfluidics
				enables rapid bacterial growth inhibition testing}},\ }\href
	{https://doi.org/10.1021/ac5001306} {\bibfield  {journal} {\bibinfo
			{journal} {Analytical Chemistry}\ }\textbf {\bibinfo {volume} {86}},\
		\bibinfo {pages} {3131} (\bibinfo {year} {2014})}\BibitemShut {NoStop}%
	\bibitem [{\citenamefont {Dehkharghani}\ \emph {et~al.}(2019)\citenamefont
		{Dehkharghani}, \citenamefont {Waisbord}, \citenamefont {Dunkel},\ and\
		\citenamefont {Guasto}}]{Dehkharghani2019}%
	\BibitemOpen
	\bibfield  {author} {\bibinfo {author} {\bibfnamefont {A.}~\bibnamefont
			{Dehkharghani}}, \bibinfo {author} {\bibfnamefont {N.}~\bibnamefont
			{Waisbord}}, \bibinfo {author} {\bibfnamefont {J.}~\bibnamefont {Dunkel}},\
		and\ \bibinfo {author} {\bibfnamefont {J.~S.}\ \bibnamefont {Guasto}},\
	}\bibfield  {title} {\bibinfo {title} {{Bacterial scattering in microfluidic
				crystal flows reveals giant active Taylor–Aris dispersion}},\ }\href
	{https://doi.org/10.1073/pnas.1819613116} {\bibfield  {journal} {\bibinfo
			{journal} {Proceedings of the National Academy of Sciences of the United
				States of America}\ }\textbf {\bibinfo {volume} {166}},\ \bibinfo {pages}
		{11119} (\bibinfo {year} {2019})}\BibitemShut {NoStop}%
	\bibitem [{\citenamefont {Narinder}\ \emph {et~al.}(2021)\citenamefont
		{Narinder}, \citenamefont {Zhu}, \citenamefont {Wei},\ and\ \citenamefont
		{Bechinger}}]{Narinder2021}%
	\BibitemOpen
	\bibfield  {author} {\bibinfo {author} {\bibfnamefont {N.}~\bibnamefont
			{Narinder}}, \bibinfo {author} {\bibfnamefont {W.-j.}\ \bibnamefont {Zhu}},
		\bibinfo {author} {\bibnamefont {Wei}},\ and\ \bibinfo {author}
		{\bibfnamefont {C.}~\bibnamefont {Bechinger}},\ }\bibfield  {title} {\bibinfo
		{title} {{Active colloids under geometrical constraints in viscoelastic
				media}},\ }\href@noop {} {\bibfield  {journal} {\bibinfo  {journal} {European
				Physical Journal E}\ }\textbf {\bibinfo {volume} {44}},\ \bibinfo {pages}
		{28} (\bibinfo {year} {2021})}\BibitemShut {NoStop}%
	\bibitem [{\citenamefont {Ishimoto}\ \emph {et~al.}(2020)\citenamefont
		{Ishimoto}, \citenamefont {Gaffney},\ and\ \citenamefont
		{Walker}}]{Ishimoto2020}%
	\BibitemOpen
	\bibfield  {author} {\bibinfo {author} {\bibfnamefont {K.}~\bibnamefont
			{Ishimoto}}, \bibinfo {author} {\bibfnamefont {E.~A.}\ \bibnamefont
			{Gaffney}},\ and\ \bibinfo {author} {\bibfnamefont {B.~J.}\ \bibnamefont
			{Walker}},\ }\bibfield  {title} {\bibinfo {title} {{Regularized
				representation of bacterial hydrodynamics}},\ }\href
	{https://doi.org/10.1103/PHYSREVFLUIDS.5.093101/FIGURES/4/MEDIUM} {\bibfield
		{journal} {\bibinfo  {journal} {Physical Review Fluids}\ }\textbf {\bibinfo
			{volume} {5}},\ \bibinfo {pages} {093101} (\bibinfo {year}
		{2020})}\BibitemShut {NoStop}%
	\bibitem [{\citenamefont {Celli}\ \emph {et~al.}(2009)\citenamefont {Celli},
		\citenamefont {Turner}, \citenamefont {Afdhal}, \citenamefont {Keates},
		\citenamefont {Ghiran}, \citenamefont {Kelly}, \citenamefont {Ewoldt},
		\citenamefont {McKinley}, \citenamefont {So}, \citenamefont {Erramilli},\
		and\ \citenamefont {Bansil}}]{Celli2009}%
	\BibitemOpen
	\bibfield  {author} {\bibinfo {author} {\bibfnamefont {J.~P.}\ \bibnamefont
			{Celli}}, \bibinfo {author} {\bibfnamefont {B.~S.}\ \bibnamefont {Turner}},
		\bibinfo {author} {\bibfnamefont {N.~H.}\ \bibnamefont {Afdhal}}, \bibinfo
		{author} {\bibfnamefont {S.}~\bibnamefont {Keates}}, \bibinfo {author}
		{\bibfnamefont {I.}~\bibnamefont {Ghiran}}, \bibinfo {author} {\bibfnamefont
			{C.~P.}\ \bibnamefont {Kelly}}, \bibinfo {author} {\bibfnamefont {R.~H.}\
			\bibnamefont {Ewoldt}}, \bibinfo {author} {\bibfnamefont {G.~H.}\
			\bibnamefont {McKinley}}, \bibinfo {author} {\bibfnamefont {P.}~\bibnamefont
			{So}}, \bibinfo {author} {\bibfnamefont {S.}~\bibnamefont {Erramilli}},\ and\
		\bibinfo {author} {\bibfnamefont {R.}~\bibnamefont {Bansil}},\ }\bibfield
	{title} {\bibinfo {title} {{Helicobacter pylori moves through mucus by
				reducing mucin viscoelasticity.}},\ }\href
	{https://doi.org/10.1073/pnas.0903438106} {\bibfield  {journal} {\bibinfo
			{journal} {Proceedings of the National Academy of Sciences of the United
				States of America}\ }\textbf {\bibinfo {volume} {106}},\ \bibinfo {pages}
		{14321} (\bibinfo {year} {2009})}\BibitemShut {NoStop}%
	\bibitem [{\citenamefont {Walker}\ \emph
		{et~al.}(2015{\natexlab{a}})\citenamefont {Walker}, \citenamefont
		{K{\"{a}}sdorf}, \citenamefont {Jeong}, \citenamefont {Lieleg},\ and\
		\citenamefont {Fischer}}]{Walker2015a}%
	\BibitemOpen
	\bibfield  {author} {\bibinfo {author} {\bibfnamefont {D.}~\bibnamefont
			{Walker}}, \bibinfo {author} {\bibfnamefont {B.~T.}\ \bibnamefont
			{K{\"{a}}sdorf}}, \bibinfo {author} {\bibfnamefont {H.~H.}\ \bibnamefont
			{Jeong}}, \bibinfo {author} {\bibfnamefont {O.}~\bibnamefont {Lieleg}},\ and\
		\bibinfo {author} {\bibfnamefont {P.}~\bibnamefont {Fischer}},\ }\bibfield
	{title} {\bibinfo {title} {{Biomolecules: Enzymatically active biomimetic
				micropropellers for the penetration of mucin gels}},\ }\href
	{https://doi.org/10.1126/sciadv.1500501} {\bibfield  {journal} {\bibinfo
			{journal} {Science Advances}\ }\textbf {\bibinfo {volume} {1}},\ \bibinfo
		{pages} {e1500501} (\bibinfo {year} {2015}{\natexlab{a}})}\BibitemShut
	{NoStop}%
	\bibitem [{\citenamefont {Guadayol}\ \emph {et~al.}(2020)\citenamefont
		{Guadayol}, \citenamefont {Mendonca}, \citenamefont {{Mariona Segura-Nogueraa
				Amand}}, \citenamefont {Wright}, \citenamefont {Tassieri},\ and\
		\citenamefont {Humphries}}]{Guadayol2020}%
	\BibitemOpen
	\bibfield  {author} {\bibinfo {author} {\bibfnamefont {{\'{O}}.}~\bibnamefont
			{Guadayol}}, \bibinfo {author} {\bibfnamefont {T.}~\bibnamefont {Mendonca}},
		\bibinfo {author} {\bibnamefont {{Mariona Segura-Nogueraa Amand}}}, \bibinfo
		{author} {\bibfnamefont {J.}~\bibnamefont {Wright}}, \bibinfo {author}
		{\bibfnamefont {M.}~\bibnamefont {Tassieri}},\ and\ \bibinfo {author}
		{\bibfnamefont {S.}~\bibnamefont {Humphries}},\ }\bibfield  {title} {\bibinfo
		{title} {{Microrheology reveals microscale viscosity gradients in planktonic
				systems}},\ }\href {https://www.pnas.org/content/118/1/e2011389118
		https://www.pnas.org/content/118/1/e2011389118.abstract} {\bibfield
		{journal} {\bibinfo  {journal} {Proceedings of the National Academy of
				Sciences of the United States of America}\ }\textbf {\bibinfo {volume} {118}}
		(\bibinfo {year} {2020})}\BibitemShut {NoStop}%
	\bibitem [{\citenamefont {Shoele}\ and\ \citenamefont
		{Eastham}(2018)}]{Shoele2018a}%
	\BibitemOpen
	\bibfield  {author} {\bibinfo {author} {\bibfnamefont {K.}~\bibnamefont
			{Shoele}}\ and\ \bibinfo {author} {\bibfnamefont {P.~S.}\ \bibnamefont
			{Eastham}},\ }\bibfield  {title} {\bibinfo {title} {{Effects of nonuniform
				viscosity on ciliary locomotion}},\ }\href
	{https://doi.org/10.1103/PhysRevFluids.3.043101} {\bibfield  {journal}
		{\bibinfo  {journal} {Physical Review Fluids}\ }\textbf {\bibinfo {volume}
			{3}},\ \bibinfo {pages} {043101} (\bibinfo {year} {2018})}\BibitemShut
	{NoStop}%
	\bibitem [{\citenamefont {Eastham}\ and\ \citenamefont
		{Shoele}(2020)}]{Eastham2020a}%
	\BibitemOpen
	\bibfield  {author} {\bibinfo {author} {\bibfnamefont {P.~S.}\ \bibnamefont
			{Eastham}}\ and\ \bibinfo {author} {\bibfnamefont {K.}~\bibnamefont
			{Shoele}},\ }\bibfield  {title} {\bibinfo {title} {{Axisymmetric squirmers in
				Stokes fluid with nonuniform viscosity}},\ }\href
	{https://doi.org/10.1103/PhysRevFluids.5.063102} {\bibfield  {journal}
		{\bibinfo  {journal} {Phys. Rev. Fluids}\ }\textbf {\bibinfo {volume} {5}},\
		\bibinfo {pages} {63102} (\bibinfo {year} {2020})}\BibitemShut {NoStop}%
	\bibitem [{\citenamefont {Dandekar}\ and\ \citenamefont
		{Ardekani}(2020)}]{Dandekar2020}%
	\BibitemOpen
	\bibfield  {author} {\bibinfo {author} {\bibfnamefont {R.}~\bibnamefont
			{Dandekar}}\ and\ \bibinfo {author} {\bibfnamefont {A.~M.}\ \bibnamefont
			{Ardekani}},\ }\bibfield  {title} {\bibinfo {title} {{Swimming sheet in a
				viscosity-stratified fluid}},\ }\href {https://doi.org/10.1017/jfm.2020.352}
	{\bibfield  {journal} {\bibinfo  {journal} {Journal of Fluid Mechanics}\
		}\textbf {\bibinfo {volume} {895}},\ \bibinfo {pages} {2} (\bibinfo {year}
		{2020})}\BibitemShut {NoStop}%
	\bibitem [{\citenamefont {Jiang}\ \emph {et~al.}(2010)\citenamefont {Jiang},
		\citenamefont {Yoshinaga},\ and\ \citenamefont {Sano}}]{Jiang2010}%
	\BibitemOpen
	\bibfield  {author} {\bibinfo {author} {\bibfnamefont {H.~R.}\ \bibnamefont
			{Jiang}}, \bibinfo {author} {\bibfnamefont {N.}~\bibnamefont {Yoshinaga}},\
		and\ \bibinfo {author} {\bibfnamefont {M.}~\bibnamefont {Sano}},\ }\bibfield
	{title} {\bibinfo {title} {{Active motion of a Janus particle by
				self-thermophoresis in a defocused laser beam}},\ }\href
	{https://journals.aps.org/prl/abstract/10.1103/PhysRevLett.105.268302}
	{\bibfield  {journal} {\bibinfo  {journal} {Physical Review Letters}\
		}\textbf {\bibinfo {volume} {105}},\ \bibinfo {pages} {268302} (\bibinfo
		{year} {2010})}\BibitemShut {NoStop}%
	\bibitem [{\citenamefont {Rings}\ \emph {et~al.}(2010)\citenamefont {Rings},
		\citenamefont {Schachoff}, \citenamefont {Selmke}, \citenamefont {Cichos},\
		and\ \citenamefont {Kroy}}]{Rings2010}%
	\BibitemOpen
	\bibfield  {author} {\bibinfo {author} {\bibfnamefont {D.}~\bibnamefont
			{Rings}}, \bibinfo {author} {\bibfnamefont {R.}~\bibnamefont {Schachoff}},
		\bibinfo {author} {\bibfnamefont {M.}~\bibnamefont {Selmke}}, \bibinfo
		{author} {\bibfnamefont {F.}~\bibnamefont {Cichos}},\ and\ \bibinfo {author}
		{\bibfnamefont {K.}~\bibnamefont {Kroy}},\ }\bibfield  {title} {\bibinfo
		{title} {{Hot Brownian motion}},\ }\href
	{https://journals.aps.org/prl/abstract/10.1103/PhysRevLett.105.090604}
	{\bibfield  {journal} {\bibinfo  {journal} {Physical Review Letters}\
		}\textbf {\bibinfo {volume} {105}},\ \bibinfo {pages} {090604} (\bibinfo
		{year} {2010})}\BibitemShut {NoStop}%
	\bibitem [{\citenamefont {Z{\"{o}}ttl}\ and\ \citenamefont
		{Yeomans}(2019)}]{Zottl2019}%
	\BibitemOpen
	\bibfield  {author} {\bibinfo {author} {\bibfnamefont {A.}~\bibnamefont
			{Z{\"{o}}ttl}}\ and\ \bibinfo {author} {\bibfnamefont {J.~M.}\ \bibnamefont
			{Yeomans}},\ }\bibfield  {title} {\bibinfo {title} {{Enhanced bacterial
				swimming speeds in macromolecular polymer solutions}},\ }\href@noop {}
	{\bibfield  {journal} {\bibinfo  {journal} {Nature Physics}\ }\textbf
		{\bibinfo {volume} {15}},\ \bibinfo {pages} {554} (\bibinfo {year}
		{2019})}\BibitemShut {NoStop}%
	\bibitem [{\citenamefont {Shen}\ \emph {et~al.}(2017)\citenamefont {Shen},
		\citenamefont {Font}, \citenamefont {Jung}, \citenamefont {Gabriel},
		\citenamefont {Stoykovich},\ and\ \citenamefont {Vernerey}}]{Shen2017}%
	\BibitemOpen
	\bibfield  {author} {\bibinfo {author} {\bibfnamefont {T.}~\bibnamefont
			{Shen}}, \bibinfo {author} {\bibfnamefont {M.~G.}\ \bibnamefont {Font}},
		\bibinfo {author} {\bibfnamefont {S.}~\bibnamefont {Jung}}, \bibinfo {author}
		{\bibfnamefont {M.~L.}\ \bibnamefont {Gabriel}}, \bibinfo {author}
		{\bibfnamefont {M.~P.}\ \bibnamefont {Stoykovich}},\ and\ \bibinfo {author}
		{\bibfnamefont {F.~J.}\ \bibnamefont {Vernerey}},\ }\bibfield  {title}
	{\bibinfo {title} {{Remotely Triggered Locomotion of Hydrogel Mag-bots in
				Confined Spaces}},\ }\href {https://doi.org/10.1038/s41598-017-16265-w}
	{\bibfield  {journal} {\bibinfo  {journal} {Scientific Reports}\ }\textbf
		{\bibinfo {volume} {7}},\ \bibinfo {pages} {1} (\bibinfo {year}
		{2017})}\BibitemShut {NoStop}%
	\bibitem [{\citenamefont {Kurzthaler}\ and\ \citenamefont
		{Stone}(2021)}]{Kurzthaler2021}%
	\BibitemOpen
	\bibfield  {author} {\bibinfo {author} {\bibfnamefont {C.}~\bibnamefont
			{Kurzthaler}}\ and\ \bibinfo {author} {\bibfnamefont {H.~A.}\ \bibnamefont
			{Stone}},\ }\bibfield  {title} {\bibinfo {title} {{Microswimmers near
				corrugated, periodic surfaces}},\ }\href {https://doi.org/10.1039/d0sm01782e}
	{\bibfield  {journal} {\bibinfo  {journal} {Soft Matter}\ }\textbf {\bibinfo
			{volume} {17}},\ \bibinfo {pages} {5} (\bibinfo {year} {2021})}\BibitemShut
	{NoStop}%
	\bibitem [{\citenamefont {Zamani}\ \emph {et~al.}(2018)\citenamefont {Zamani},
		\citenamefont {Ebrahimtabar}, \citenamefont {Zamani}, \citenamefont {Miller},
		\citenamefont {Alizadeh-Navaei}, \citenamefont {Shokri-Shirvani},\ and\
		\citenamefont {Derakhshan}}]{Zamani2018}%
	\BibitemOpen
	\bibfield  {author} {\bibinfo {author} {\bibfnamefont {M.}~\bibnamefont
			{Zamani}}, \bibinfo {author} {\bibfnamefont {F.}~\bibnamefont
			{Ebrahimtabar}}, \bibinfo {author} {\bibfnamefont {V.}~\bibnamefont
			{Zamani}}, \bibinfo {author} {\bibfnamefont {W.~H.}\ \bibnamefont {Miller}},
		\bibinfo {author} {\bibfnamefont {R.}~\bibnamefont {Alizadeh-Navaei}},
		\bibinfo {author} {\bibfnamefont {J.}~\bibnamefont {Shokri-Shirvani}},\ and\
		\bibinfo {author} {\bibfnamefont {M.~H.}\ \bibnamefont {Derakhshan}},\
	}\bibfield  {title} {\bibinfo {title} {{Systematic review with meta-analysis:
				the worldwide prevalence of {\textless}i{\textgreater}Helicobacter
				pylori{\textless}/i{\textgreater} infection}},\ }\href
	{https://doi.org/10.1111/apt.14561} {\bibfield  {journal} {\bibinfo
			{journal} {Alimentary Pharmacology {\&} Therapeutics}\ }\textbf {\bibinfo
			{volume} {47}},\ \bibinfo {pages} {868} (\bibinfo {year} {2018})}\BibitemShut
	{NoStop}%
	\bibitem [{\citenamefont {Celli}\ \emph {et~al.}(2007)\citenamefont {Celli},
		\citenamefont {Turner}, \citenamefont {Afdhal}, \citenamefont {Ewoldt},
		\citenamefont {McKinley}, \citenamefont {Bansil},\ and\ \citenamefont
		{Erramilli}}]{Celli2007}%
	\BibitemOpen
	\bibfield  {author} {\bibinfo {author} {\bibfnamefont {J.~P.}\ \bibnamefont
			{Celli}}, \bibinfo {author} {\bibfnamefont {B.~S.}\ \bibnamefont {Turner}},
		\bibinfo {author} {\bibfnamefont {N.~H.}\ \bibnamefont {Afdhal}}, \bibinfo
		{author} {\bibfnamefont {R.~H.}\ \bibnamefont {Ewoldt}}, \bibinfo {author}
		{\bibfnamefont {G.~H.}\ \bibnamefont {McKinley}}, \bibinfo {author}
		{\bibfnamefont {R.}~\bibnamefont {Bansil}},\ and\ \bibinfo {author}
		{\bibfnamefont {S.}~\bibnamefont {Erramilli}},\ }\bibfield  {title} {\bibinfo
		{title} {{Rheology of gastric mucin exhibits a pH-dependent sol-gel
				transition}},\ }\href {https://doi.org/10.1021/bm0609691} {\bibfield
		{journal} {\bibinfo  {journal} {Biomacromolecules}\ }\textbf {\bibinfo
			{volume} {8}},\ \bibinfo {pages} {1580} (\bibinfo {year} {2007})}\BibitemShut
	{NoStop}%
	\bibitem [{\citenamefont {Mirbagheri}\ and\ \citenamefont
		{Fu}(2016)}]{Mirbagheri2016}%
	\BibitemOpen
	\bibfield  {author} {\bibinfo {author} {\bibfnamefont {S.~A.}\ \bibnamefont
			{Mirbagheri}}\ and\ \bibinfo {author} {\bibfnamefont {H.~C.}\ \bibnamefont
			{Fu}},\ }\bibfield  {title} {\bibinfo {title} {{Helicobacter pylori Couples
				Motility and Diffusion to Actively Create a Heterogeneous Complex Medium in
				Gastric Mucus}},\ }\href {https://doi.org/10.1103/PhysRevLett.116.198101}
	{\bibfield  {journal} {\bibinfo  {journal} {Physical Review Letters}\
		}\textbf {\bibinfo {volume} {116}},\ \bibinfo {pages} {198101} (\bibinfo
		{year} {2016})}\BibitemShut {NoStop}%
	\bibitem [{\citenamefont {Leshansky}(2009)}]{Leshansky2009}%
	\BibitemOpen
	\bibfield  {author} {\bibinfo {author} {\bibfnamefont {A.~M.}\ \bibnamefont
			{Leshansky}},\ }\bibfield  {title} {\bibinfo {title} {{Enhanced
				low-Reynolds-number propulsion in heterogeneous viscous environments}},\
	}\href@noop {} {\bibfield  {journal} {\bibinfo  {journal} {Physical Review
				E}\ }\textbf {\bibinfo {volume} {80}} (\bibinfo {year} {2009})}\BibitemShut
	{NoStop}%
	\bibitem [{\citenamefont {Reigh}\ and\ \citenamefont
		{Lauga}(2017)}]{Reigh2017}%
	\BibitemOpen
	\bibfield  {author} {\bibinfo {author} {\bibfnamefont {S.~Y.}\ \bibnamefont
			{Reigh}}\ and\ \bibinfo {author} {\bibfnamefont {E.}~\bibnamefont {Lauga}},\
	}\bibfield  {title} {\bibinfo {title} {{Two-fluid model for locomotion under
				self-confinement}},\ }\href {https://doi.org/10.1103/PhysRevFluids.2.093101}
	{\bibfield  {journal} {\bibinfo  {journal} {Physical Review Fluids}\ }\textbf
		{\bibinfo {volume} {2}},\ \bibinfo {pages} {093101} (\bibinfo {year}
		{2017})}\BibitemShut {NoStop}%
	\bibitem [{\citenamefont {Nganguia}\ \emph {et~al.}(2020)\citenamefont
		{Nganguia}, \citenamefont {Zhu}, \citenamefont {Palaniappan},\ and\
		\citenamefont {Pak}}]{Nganguia2020}%
	\BibitemOpen
	\bibfield  {author} {\bibinfo {author} {\bibfnamefont {H.}~\bibnamefont
			{Nganguia}}, \bibinfo {author} {\bibfnamefont {L.}~\bibnamefont {Zhu}},
		\bibinfo {author} {\bibfnamefont {D.}~\bibnamefont {Palaniappan}},\ and\
		\bibinfo {author} {\bibfnamefont {O.~S.}\ \bibnamefont {Pak}},\ }\bibfield
	{title} {\bibinfo {title} {{Squirming in a viscous fluid enclosed by a
				Brinkman medium}},\ }\href {https://doi.org/10.1103/PhysRevE.101.063105}
	{\bibfield  {journal} {\bibinfo  {journal} {Physical Review E}\ }\textbf
		{\bibinfo {volume} {101}},\ \bibinfo {pages} {063105} (\bibinfo {year}
		{2020})}\BibitemShut {NoStop}%
	\bibitem [{\citenamefont {Walker}\ \emph
		{et~al.}(2015{\natexlab{b}})\citenamefont {Walker}, \citenamefont
		{K{\"{a}}sdorf}, \citenamefont {Jeong}, \citenamefont {Lieleg},\ and\
		\citenamefont {Fischer}}]{Walker2015}%
	\BibitemOpen
	\bibfield  {author} {\bibinfo {author} {\bibfnamefont {D.}~\bibnamefont
			{Walker}}, \bibinfo {author} {\bibfnamefont {B.~T.}\ \bibnamefont
			{K{\"{a}}sdorf}}, \bibinfo {author} {\bibfnamefont {H.-H.}\ \bibnamefont
			{Jeong}}, \bibinfo {author} {\bibfnamefont {O.}~\bibnamefont {Lieleg}},\ and\
		\bibinfo {author} {\bibfnamefont {P.}~\bibnamefont {Fischer}},\ }\bibfield
	{title} {\bibinfo {title} {{Enzymatically active biomimetic micropropellers
				for the penetration of mucin gels}},\ }\href
	{https://doi.org/10.1126/sciadv.1500501} {\bibfield  {journal} {\bibinfo
			{journal} {Science Advances}\ }\textbf {\bibinfo {volume} {1}},\ \bibinfo
		{pages} {e1500501} (\bibinfo {year} {2015}{\natexlab{b}})}\BibitemShut
	{NoStop}%
	\bibitem [{\citenamefont {Cortez}(2002)}]{Cortez2002}%
	\BibitemOpen
	\bibfield  {author} {\bibinfo {author} {\bibfnamefont {R.}~\bibnamefont
			{Cortez}},\ }\bibfield  {title} {\bibinfo {title} {{The method of regularized
				stokeslets}},\ }\href {https://doi.org/10.1137/S106482750038146X} {\bibfield
		{journal} {\bibinfo  {journal} {SIAM Journal on Scientific Computing}\
		}\textbf {\bibinfo {volume} {23}},\ \bibinfo {pages} {1204} (\bibinfo {year}
		{2002})}\BibitemShut {NoStop}%
	\bibitem [{\citenamefont {Cortez}\ \emph {et~al.}(2005)\citenamefont {Cortez},
		\citenamefont {Fauci},\ and\ \citenamefont {Medovikov}}]{Cortez2005}%
	\BibitemOpen
	\bibfield  {author} {\bibinfo {author} {\bibfnamefont {R.}~\bibnamefont
			{Cortez}}, \bibinfo {author} {\bibfnamefont {L.}~\bibnamefont {Fauci}},\ and\
		\bibinfo {author} {\bibfnamefont {A.}~\bibnamefont {Medovikov}},\ }\bibfield
	{title} {\bibinfo {title} {{The method of regularized Stokeslets in three
				dimensions: Analysis, validation, and application to helical swimming}},\
	}\href {https://doi.org/10.1063/1.1830486} {\bibfield  {journal} {\bibinfo
			{journal} {Physics of Fluids}\ }\textbf {\bibinfo {volume} {17}},\ \bibinfo
		{pages} {031504} (\bibinfo {year} {2005})}\BibitemShut {NoStop}%
	\bibitem [{\citenamefont {Hyon}\ \emph {et~al.}(2012)\citenamefont {Hyon},
		\citenamefont {Marcos}, \citenamefont {Powers}, \citenamefont {Stocker},\
		and\ \citenamefont {Fu}}]{Hyon2012}%
	\BibitemOpen
	\bibfield  {author} {\bibinfo {author} {\bibfnamefont {Y.}~\bibnamefont
			{Hyon}}, \bibinfo {author} {\bibnamefont {Marcos}}, \bibinfo {author}
		{\bibfnamefont {T.~R.}\ \bibnamefont {Powers}}, \bibinfo {author}
		{\bibfnamefont {R.}~\bibnamefont {Stocker}},\ and\ \bibinfo {author}
		{\bibfnamefont {H.~C.}\ \bibnamefont {Fu}},\ }\bibfield  {title} {\bibinfo
		{title} {{The wiggling trajectories of bacteria}},\ }\href
	{https://doi.org/10.1017/jfm.2012.217} {\bibfield  {journal} {\bibinfo
			{journal} {Journal of Fluid Mechanics}\ }\textbf {\bibinfo {volume} {705}},\
		\bibinfo {pages} {58} (\bibinfo {year} {2012})}\BibitemShut {NoStop}%
	\bibitem [{\citenamefont {Martindale}\ \emph {et~al.}(2016)\citenamefont
		{Martindale}, \citenamefont {Jabbarzadeh},\ and\ \citenamefont
		{Fu}}]{Martindale2016}%
	\BibitemOpen
	\bibfield  {author} {\bibinfo {author} {\bibfnamefont {J.~D.}\ \bibnamefont
			{Martindale}}, \bibinfo {author} {\bibfnamefont {M.}~\bibnamefont
			{Jabbarzadeh}},\ and\ \bibinfo {author} {\bibfnamefont {H.~C.}\ \bibnamefont
			{Fu}},\ }\bibfield  {title} {\bibinfo {title} {{Choice of computational
				method for swimming and pumping with nonslender helical filaments at low
				Reynolds number}},\ }\href {https://doi.org/10.1063/1.4940904} {\bibfield
		{journal} {\bibinfo  {journal} {Physics of Fluids}\ }\textbf {\bibinfo
			{volume} {28}},\ \bibinfo {pages} {021901} (\bibinfo {year}
		{2016})}\BibitemShut {NoStop}%
	\bibitem [{\citenamefont {Kamarapu}\ \emph {et~al.}(2021)\citenamefont
		{Kamarapu}, \citenamefont {Jabbarzadeh},\ and\ \citenamefont
		{Fu}}]{Kamarapu2021}%
	\BibitemOpen
	\bibfield  {author} {\bibinfo {author} {\bibfnamefont {S.~K.}\ \bibnamefont
			{Kamarapu}}, \bibinfo {author} {\bibfnamefont {M.}~\bibnamefont
			{Jabbarzadeh}},\ and\ \bibinfo {author} {\bibfnamefont {H.~C.}\ \bibnamefont
			{Fu}},\ }\bibfield  {title} {\bibinfo {title} {{Modeling creeping flows in
				porous media using regularized Stokeslets}},\ }\href
	{https://arxiv.org/abs/2110.05548v1} {\  (\bibinfo {year} {2021})},\ \Eprint
	{https://arxiv.org/abs/2110.05548} {arXiv:2110.05548} \BibitemShut {NoStop}%
	\bibitem [{\citenamefont {Rostami}\ and\ \citenamefont
		{Olson}(2016)}]{Rostami2016}%
	\BibitemOpen
	\bibfield  {author} {\bibinfo {author} {\bibfnamefont {M.~W.}\ \bibnamefont
			{Rostami}}\ and\ \bibinfo {author} {\bibfnamefont {S.~D.}\ \bibnamefont
			{Olson}},\ }\bibfield  {title} {\bibinfo {title} {{Kernel-independent fast
				multipole method within the framework of regularized Stokeslets}},\ }\href
	{https://doi.org/10.1016/j.jfluidstructs.2016.07.006} {\bibfield  {journal}
		{\bibinfo  {journal} {Journal of Fluids and Structures}\ }\textbf {\bibinfo
			{volume} {67}},\ \bibinfo {pages} {60} (\bibinfo {year} {2016})}\BibitemShut
	{NoStop}%
	\bibitem [{\citenamefont {Rostami}\ and\ \citenamefont
		{Olson}(2019)}]{Rostami2019}%
	\BibitemOpen
	\bibfield  {author} {\bibinfo {author} {\bibfnamefont {M.~W.}\ \bibnamefont
			{Rostami}}\ and\ \bibinfo {author} {\bibfnamefont {S.~D.}\ \bibnamefont
			{Olson}},\ }\bibfield  {title} {\bibinfo {title} {{Fast algorithms for large
				dense matrices with applications to biofluids}},\ }\href
	{https://doi.org/10.1016/j.jcp.2019.05.042} {\bibfield  {journal} {\bibinfo
			{journal} {Journal of Computational Physics}\ }\textbf {\bibinfo {volume}
			{394}},\ \bibinfo {pages} {364} (\bibinfo {year} {2019})}\BibitemShut
	{NoStop}%
	\bibitem [{\citenamefont {Mart{\'{i}}nez}\ \emph {et~al.}(2016)\citenamefont
		{Mart{\'{i}}nez}, \citenamefont {Hardcastle}, \citenamefont {Wang},
		\citenamefont {Pincus}, \citenamefont {Tsang}, \citenamefont {Hoover},
		\citenamefont {Bansil},\ and\ \citenamefont {Salama}}]{Martinez2016}%
	\BibitemOpen
	\bibfield  {author} {\bibinfo {author} {\bibfnamefont {L.~E.}\ \bibnamefont
			{Mart{\'{i}}nez}}, \bibinfo {author} {\bibfnamefont {J.~M.}\ \bibnamefont
			{Hardcastle}}, \bibinfo {author} {\bibfnamefont {J.}~\bibnamefont {Wang}},
		\bibinfo {author} {\bibfnamefont {Z.}~\bibnamefont {Pincus}}, \bibinfo
		{author} {\bibfnamefont {J.}~\bibnamefont {Tsang}}, \bibinfo {author}
		{\bibfnamefont {T.~R.}\ \bibnamefont {Hoover}}, \bibinfo {author}
		{\bibfnamefont {R.}~\bibnamefont {Bansil}},\ and\ \bibinfo {author}
		{\bibfnamefont {N.~R.}\ \bibnamefont {Salama}},\ }\bibfield  {title}
	{\bibinfo {title} {{Helicobacter pylori strains vary cell shape and flagellum
				number to maintain robust motility in viscous environments}},\ }\href
	{https://doi.org/10.1111/mmi.13218} {\bibfield  {journal} {\bibinfo
			{journal} {Molecular Microbiology}\ }\textbf {\bibinfo {volume} {99}},\
		\bibinfo {pages} {88} (\bibinfo {year} {2016})}\BibitemShut {NoStop}%
	\bibitem [{\citenamefont {{E. L. Cussler}}(2009)}]{Cussler}%
	\BibitemOpen
	\bibfield  {author} {\bibinfo {author} {\bibnamefont {{E. L. Cussler}}},\
	}\href
	{https://www.cambridge.org/us/academic/subjects/engineering/chemical-engineering/diffusion-mass-transfer-fluid-systems-3rd-edition?format=HB{\&}isbn=9780521871211}
	{\emph {\bibinfo {title} {{Diffusion mass transfer fluid systems}}}},\
	\bibinfo {edition} {3rd}\ ed.\ (\bibinfo  {publisher} {Cambridge University
		Press},\ \bibinfo {year} {2009})\BibitemShut {NoStop}%
	\bibitem [{\citenamefont {Fu}\ \emph {et~al.}(2010)\citenamefont {Fu},
		\citenamefont {Shenoy},\ and\ \citenamefont {Powers}}]{Fu2010a}%
	\BibitemOpen
	\bibfield  {author} {\bibinfo {author} {\bibfnamefont {H.~C.}\ \bibnamefont
			{Fu}}, \bibinfo {author} {\bibfnamefont {V.~B.}\ \bibnamefont {Shenoy}},\
		and\ \bibinfo {author} {\bibfnamefont {T.~R.}\ \bibnamefont {Powers}},\
	}\bibfield  {title} {\bibinfo {title} {{Low-Reynolds-number swimming in
				gels}},\ }\href {https://doi.org/10.1209/0295-5075/91/24002} {\bibfield
		{journal} {\bibinfo  {journal} {EPL (Europhysics Letters)}\ }\textbf
		{\bibinfo {volume} {91}},\ \bibinfo {pages} {24002} (\bibinfo {year}
		{2010})}\BibitemShut {NoStop}%
	\bibitem [{\citenamefont {Purcell}(1977)}]{Purcell1977}%
	\BibitemOpen
	\bibfield  {author} {\bibinfo {author} {\bibfnamefont {E.~M.}\ \bibnamefont
			{Purcell}},\ }\bibfield  {title} {\bibinfo {title} {{Life at low Reynolds
				number}},\ }\href {https://doi.org/10.1119/1.10903} {\bibfield  {journal}
		{\bibinfo  {journal} {American Journal of Physics}\ }\textbf {\bibinfo
			{volume} {45}},\ \bibinfo {pages} {3} (\bibinfo {year} {1977})}\BibitemShut
	{NoStop}%
	\bibitem [{\citenamefont {Jabbarzadeh}\ and\ \citenamefont
		{Fu}(2018)}]{Jabbarzadeh2018}%
	\BibitemOpen
	\bibfield  {author} {\bibinfo {author} {\bibfnamefont {M.}~\bibnamefont
			{Jabbarzadeh}}\ and\ \bibinfo {author} {\bibfnamefont {H.~C.}\ \bibnamefont
			{Fu}},\ }\bibfield  {title} {\bibinfo {title} {Viscous constraints on
			microorganism approach and interaction},\ }\href@noop {} {\bibfield
		{journal} {\bibinfo  {journal} {J. Fluid Mechanics}\ }\textbf {\bibinfo
			{volume} {851}},\ \bibinfo {pages} {715} (\bibinfo {year}
		{2018})}\BibitemShut {NoStop}%
	\bibitem [{\citenamefont {Qu}\ and\ \citenamefont {Breuer}(2020)}]{Qu2020}%
	\BibitemOpen
	\bibfield  {author} {\bibinfo {author} {\bibfnamefont {Z.}~\bibnamefont
			{Qu}}\ and\ \bibinfo {author} {\bibfnamefont {K.~S.}\ \bibnamefont
			{Breuer}},\ }\bibfield  {title} {\bibinfo {title} {{Effects of shear-thinning
				viscosity and viscoelastic stresses on flagellated bacteria motility}},\
	}\href
	{https://journals.aps.org/prfluids/abstract/10.1103/PhysRevFluids.5.073103}
	{\bibfield  {journal} {\bibinfo  {journal} {Physical Review Fluids}\ }\textbf
		{\bibinfo {volume} {5}},\ \bibinfo {pages} {073103} (\bibinfo {year}
		{2020})}\BibitemShut {NoStop}%
	\bibitem [{\citenamefont {Li}\ and\ \citenamefont {Ardekani}(2015)}]{Li2015}%
	\BibitemOpen
	\bibfield  {author} {\bibinfo {author} {\bibfnamefont {G.}~\bibnamefont
			{Li}}\ and\ \bibinfo {author} {\bibfnamefont {A.~M.}\ \bibnamefont
			{Ardekani}},\ }\bibfield  {title} {\bibinfo {title} {{Undulatory swimming in
				non-Newtonian fluids}},\ }\href {https://doi.org/10.1017/jfm.2015.595}
	{\bibfield  {journal} {\bibinfo  {journal} {Journal of Fluid Mechanics}\
		}\textbf {\bibinfo {volume} {784}},\ \bibinfo {pages} {R4} (\bibinfo {year}
		{2015})}\BibitemShut {NoStop}%
	\bibitem [{\citenamefont {G{\'{o}}mez}\ \emph {et~al.}(2017)\citenamefont
		{G{\'{o}}mez}, \citenamefont {God{\'{i}}nez}, \citenamefont {Lauga},\ and\
		\citenamefont {Zenit}}]{Gomez2017}%
	\BibitemOpen
	\bibfield  {author} {\bibinfo {author} {\bibfnamefont {S.}~\bibnamefont
			{G{\'{o}}mez}}, \bibinfo {author} {\bibfnamefont {F.~A.}\ \bibnamefont
			{God{\'{i}}nez}}, \bibinfo {author} {\bibfnamefont {E.}~\bibnamefont
			{Lauga}},\ and\ \bibinfo {author} {\bibfnamefont {R.}~\bibnamefont {Zenit}},\
	}\bibfield  {title} {\bibinfo {title} {{Helical propulsion in shear-thinning
				fluids}},\ }\href {https://doi.org/10.1017/jfm.2016.807} {\bibfield
		{journal} {\bibinfo  {journal} {Journal of Fluid Mechanics}\ }\textbf
		{\bibinfo {volume} {812}},\ \bibinfo {pages} {R3} (\bibinfo {year}
		{2017})}\BibitemShut {NoStop}%
	\bibitem [{\citenamefont {Pietrzyk}\ \emph {et~al.}(2019)\citenamefont
		{Pietrzyk}, \citenamefont {Nganguia}, \citenamefont {Datt}, \citenamefont
		{Zhu}, \citenamefont {Elfring},\ and\ \citenamefont {Pak}}]{Pietrzyk2019}%
	\BibitemOpen
	\bibfield  {author} {\bibinfo {author} {\bibfnamefont {K.}~\bibnamefont
			{Pietrzyk}}, \bibinfo {author} {\bibfnamefont {H.}~\bibnamefont {Nganguia}},
		\bibinfo {author} {\bibfnamefont {C.}~\bibnamefont {Datt}}, \bibinfo {author}
		{\bibfnamefont {L.}~\bibnamefont {Zhu}}, \bibinfo {author} {\bibfnamefont
			{G.~J.}\ \bibnamefont {Elfring}},\ and\ \bibinfo {author} {\bibfnamefont
			{O.~S.}\ \bibnamefont {Pak}},\ }\bibfield  {title} {\bibinfo {title} {{Flow
				around a squirmer in a shear-thinning fluid}},\ }\href
	{https://doi.org/10.1016/j.jnnfm.2019.04.005} {\bibfield  {journal} {\bibinfo
			{journal} {Journal of Non-Newtonian Fluid Mechanics}\ }\textbf {\bibinfo
			{volume} {268}},\ \bibinfo {pages} {101} (\bibinfo {year}
		{2019})}\BibitemShut {NoStop}%
	\bibitem [{\citenamefont {Hewitt}\ and\ \citenamefont
		{Balmforth}(2017)}]{Hewitt2017}%
	\BibitemOpen
	\bibfield  {author} {\bibinfo {author} {\bibfnamefont {D.~R.}\ \bibnamefont
			{Hewitt}}\ and\ \bibinfo {author} {\bibfnamefont {N.~J.}\ \bibnamefont
			{Balmforth}},\ }\bibfield  {title} {\bibinfo {title} {{Taylor's swimming
				sheet in a yield-stress fluid}},\ }\href
	{https://doi.org/10.1017/jfm.2017.476} {\bibfield  {journal} {\bibinfo
			{journal} {Journal of Fluid Mechanics}\ }\textbf {\bibinfo {volume} {828}},\
		\bibinfo {pages} {33} (\bibinfo {year} {2017})}\BibitemShut {NoStop}%
	\bibitem [{\citenamefont {{Esparza L{\'{o}}pez}}\ \emph
		{et~al.}(2021)\citenamefont {{Esparza L{\'{o}}pez}}, \citenamefont
		{Gonzalez-Gutierrez}, \citenamefont {Solorio-Ordaz}, \citenamefont {Lauga},\
		and\ \citenamefont {Zenit}}]{EsparzaLopez2021}%
	\BibitemOpen
	\bibfield  {author} {\bibinfo {author} {\bibfnamefont {C.}~\bibnamefont
			{{Esparza L{\'{o}}pez}}}, \bibinfo {author} {\bibfnamefont {J.}~\bibnamefont
			{Gonzalez-Gutierrez}}, \bibinfo {author} {\bibfnamefont {F.}~\bibnamefont
			{Solorio-Ordaz}}, \bibinfo {author} {\bibfnamefont {E.}~\bibnamefont
			{Lauga}},\ and\ \bibinfo {author} {\bibfnamefont {R.}~\bibnamefont {Zenit}},\
	}\bibfield  {title} {\bibinfo {title} {{Dynamics of a helical swimmer
				crossing viscosity gradients}},\ }\href@noop {} {\bibfield  {journal}
		{\bibinfo  {journal} {Physical Review Fluids}\ }\textbf {\bibinfo {volume}
			{6}},\ \bibinfo {pages} {083102} (\bibinfo {year} {2021})}\BibitemShut
	{NoStop}%
	\bibitem [{\citenamefont {Meshkati}\ and\ \citenamefont
		{Fu}(2014)}]{Meshkati2014}%
	\BibitemOpen
	\bibfield  {author} {\bibinfo {author} {\bibfnamefont {F.}~\bibnamefont
			{Meshkati}}\ and\ \bibinfo {author} {\bibfnamefont {H.~C.}\ \bibnamefont
			{Fu}},\ }\bibfield  {title} {\bibinfo {title} {{Modeling rigid magnetically
				rotated microswimmers: Rotation axes, bistability, and controllability}},\
	}\href {https://doi.org/10.1103/PhysRevE.90.063006} {\bibfield  {journal}
		{\bibinfo  {journal} {Physical Review E - Statistical, Nonlinear, and Soft
				Matter Physics}\ }\textbf {\bibinfo {volume} {90}},\ \bibinfo {pages}
		{063006} (\bibinfo {year} {2014})}\BibitemShut {NoStop}%
	\bibitem [{\citenamefont {Saad}\ and\ \citenamefont
		{Schultz}(1986)}]{Saad1986}%
	\BibitemOpen
	\bibfield  {author} {\bibinfo {author} {\bibfnamefont {Y.}~\bibnamefont
			{Saad}}\ and\ \bibinfo {author} {\bibfnamefont {M.~H.}\ \bibnamefont
			{Schultz}},\ }\bibfield  {title} {\bibinfo {title} {{GMRES: A Generalized
				Minimal Residual Algorithm for Solving Nonsymmetric Linear Systems}},\ }\href
	{https://doi.org/10.1137/0907058} {\bibfield  {journal} {\bibinfo  {journal}
			{SIAM Journal on Scientific and Statistical Computing}\ }\textbf {\bibinfo
			{volume} {7}},\ \bibinfo {pages} {856} (\bibinfo {year} {1986})}\BibitemShut
	{NoStop}%
	\bibitem [{\citenamefont {Constantino}\ \emph {et~al.}(2016)\citenamefont
		{Constantino}, \citenamefont {Jabbarzadeh}, \citenamefont {Fu},\ and\
		\citenamefont {Bansil}}]{Constantino2016}%
	\BibitemOpen
	\bibfield  {author} {\bibinfo {author} {\bibfnamefont {M.~A.}\ \bibnamefont
			{Constantino}}, \bibinfo {author} {\bibfnamefont {M.}~\bibnamefont
			{Jabbarzadeh}}, \bibinfo {author} {\bibfnamefont {H.~C.}\ \bibnamefont
			{Fu}},\ and\ \bibinfo {author} {\bibfnamefont {R.}~\bibnamefont {Bansil}},\
	}\bibfield  {title} {\bibinfo {title} {{Helical and rod-shaped bacteria swim
				in helical trajectories with little additional propulsion from helical
				shape}},\ }\href {https://doi.org/10.1126/sciadv.1601661} {\bibfield
		{journal} {\bibinfo  {journal} {Science Advances}\ }\textbf {\bibinfo
			{volume} {2}},\ \bibinfo {pages} {e1601661} (\bibinfo {year}
		{2016})}\BibitemShut {NoStop}%
	\bibitem [{\citenamefont {Durlofsky}\ and\ \citenamefont
		{Brady}(1987)}]{Durlofsky1987}%
	\BibitemOpen
	\bibfield  {author} {\bibinfo {author} {\bibfnamefont {L.}~\bibnamefont
			{Durlofsky}}\ and\ \bibinfo {author} {\bibfnamefont {J.~F.}\ \bibnamefont
			{Brady}},\ }\bibfield  {title} {\bibinfo {title} {{Analysis of the Brinkman
				equation as a model for flow in porous media.}},\ }\href
	{https://doi.org/10.1063/1.866465} {\bibfield  {journal} {\bibinfo  {journal}
			{PHYS. FLUIDS}\ }\textbf {\bibinfo {volume} {30}},\ \bibinfo {pages} {3329}
		(\bibinfo {year} {1987})}\BibitemShut {NoStop}%
	\bibitem [{\citenamefont {Happel}\ and\ \citenamefont
		{Brenner}(1983)}]{Happel1983}%
	\BibitemOpen
	\bibfield  {author} {\bibinfo {author} {\bibfnamefont {J.}~\bibnamefont
			{Happel}}\ and\ \bibinfo {author} {\bibfnamefont {H.}~\bibnamefont
			{Brenner}},\ }\bibfield  {title} {\bibinfo {title} {{Axisymmetrical Flow}}\
	}(\bibinfo {year} {1983})\ pp.\ \bibinfo {pages} {96--158}\BibitemShut
	{NoStop}%
	\bibitem [{\citenamefont {Stimson}\ and\ \citenamefont
		{Jeffery}(1926)}]{Stimson1926}%
	\BibitemOpen
	\bibfield  {author} {\bibinfo {author} {\bibfnamefont {M.}~\bibnamefont
			{Stimson}}\ and\ \bibinfo {author} {\bibfnamefont {G.}~\bibnamefont
			{Jeffery}},\ }\bibfield  {title} {\bibinfo {title} {The motion of two spheres
			in a viscous fluid},\ }\href@noop {} {\bibfield  {journal} {\bibinfo
			{journal} {Proceedings of the Royal Society of London. Series A, Containing
				Papers of a Mathematical and Physical Character}\ }\textbf {\bibinfo {volume}
			{111}},\ \bibinfo {pages} {110} (\bibinfo {year} {1926})}\BibitemShut
	{NoStop}%
\end{thebibliography}
%

\end{document}